\documentclass[twocolumn,amssymb, nobibnotes, aps, prd,amsmath,mathtools, nofootinbib,pgfmath,pgffor,superscriptaddress]{revtex4-1}
\usepackage{graphicx}
\usepackage{ulem,amsthm}

\newcommand{\nn}{\nonumber}
\newcommand{\be}{\begin{equation}}
\newcommand{\ee}{\end{equation}}
\makeatletter
\def\qrr@split@result#1 #2\@qrr@split@result{\edef\erfInput{#1}\edef\erfResult{#2}}
\newcommand*{\gnuplotErf}[2][\jobname.eval]{%
    \immediate\write18{gnuplot -e "set print '#1'; print #2, erf(#2);"}%
    \everyeof{\noexpand}
    \edef\qrr@temp{\input{#1} }%
    \expandafter\qrr@split@result\qrr@temp\@qrr@split@result
}
\makeatother

\usepackage{caption,color}
\usepackage{subcaption}
\usepackage{multirow}

\usepackage{hyperref}
\hypersetup{colorlinks,citecolor=blue}

\usepackage[dvipsnames]{xcolor}
\usepackage{color, colortbl}
\usepackage{bigints}

\DeclareUnicodeCharacter{0301}{í}

\begin{document}

\title[GW190521: First Measurement of Stimulated Hawking Radiation from Black Holes]{GW190521: Search for Echoes due to Stimulated Hawking Radiation from Black Holes}
\author{Jahed Abedi}
\email{jahed.abedi@uis.no}
\affiliation{Department of Mathematics and Physics, University of Stavanger, NO-4036 Stavanger, Norway}
\affiliation{Max-Planck-Institut f\"ur Gravitationsphysik, D-30167 Hannover, Germany}
\affiliation{Leibniz Universit\"at Hannover, D-30167 Hannover, Germany}

\author{Luís Felipe Longo Micchi}
\email{luis.longo@ufabc.edu.br}
\affiliation{Center for Natural and Human Sciences, UFABC, Santo Andŕe, SP 09210-170, Brazil}

\author{Niayesh Afshordi}
\email[]{nafshordi@pitp.ca }
\affiliation{Department of Physics and Astronomy, University of Waterloo,
200 University Ave W, N2L 3G1, Waterloo, Canada
}
\affiliation{Waterloo Centre for Astrophysics, University of Waterloo, Waterloo, ON, N2L 3G1, Canada}
\affiliation{Perimeter Institute For Theoretical Physics, 31 Caroline St N, Waterloo, Canada}

\begin{abstract}
Being arguably the most massive binary black hole merger event observed to date, GW190521 deserves special attention. The exceptionally loud ringdown of this merger makes it an ideal candidate to search for gravitational wave echoes, a proposed smoking gun for the quantum structure of black hole horizons. We perform an unprecedented multi-pronged search for echoes via two well-established and independent pipelines: a template-based search for stimulated emission of Hawking radiation, or Boltzmann echoes, and the model-agnostic coherent WaveBurst (cWB) search. Stimulated Hawking radiation from the merger is expected to lead to post-merger echoes at horizon mode frequency of $\sim 50$ Hz (for quadrupolar gravitational radiation), repeating at intervals of $\sim 1$ second, due to partial reflection off Planckian quantum structure of the horizon. A careful analysis using dynamic nested sampling yields a Bayesian evidence of $8^{+4}_{-2}$ (90\% confidence level) for this signal following GW190521, carrying an excess of $6^{+10}_{-5}\%$ in gravitational wave energy, relative to the main event (consistent with the predicted amplitude of Boltzmann echoes). The ``look-elsewhere'' effect is estimated by using General Relativity (plus Boltzmann echoes) injections in real data, before and after the event, giving a false (true) positive detection probability for higher Bayes factors of $1.5^{+1.2}_{-0.9} \%$  ($35 \pm 7 \%$). Similarly, the reconstructed waveform of the first echo in cWB carries an energy excess of $13^{+16}_{-7}\%$. While the current evidence for stimulated Hawking radiation does not reach the gold standard of $5\sigma$ (or p-value  $< 3 \times 10^{-7}$), our findings are in line with expectations for stimulated Hawking radiation at current detector sensitivities. The next generation of gravitational wave observatories can thus draw a definitive conclusion on the quantum nature of black hole horizons. 
\end{abstract}

\maketitle


\section{Introduction}
The rise of gravitational wave (GW) astronomy has opened an unprecedented window into the mysterious nature of quantum black holes (BHs) that has fascinated theoretical physicists for nearly half a century \cite{Cardoso:2019rvt,Abedi:2020ujo}. A smoking gun for the quantum structure of BH horizons, motivated by the information paradox \cite{Almheiri:2012rt,Polchinski:2016hrw}, is the potential appearance of GW echoes, repeating at $0.1-1$ second intervals, that may follow detection of compact binary merger events that involve stellar BHs \cite{Cardoso:2016rao,Cardoso:2016oxy}. One may consider echoes as stimulated emission of Hawking radiation, caused by the GWs that excite the quantum BH microstructure \cite{Oshita:2019sat,Wang:2019rcf,Xin:2021zir,Srivastava:2021uku}.  While stimulated Hawking radiation has been seen in analogue BH/white hole systems, based on water waves, nonlinear optics, or Bose-Einstein condensates, for nearly a decade \cite{Weinfurtner:2010nu,Steinhauer:2014dra,Wang:2016jaj,Drori:2018ivu,Bera:2020doh,Steinhauer:2021xxj}, they have remained elusive in real gravitational BHs.

An advantage of testing General Relativity (GR) using echoes is that while modeling the strong-field regime of mergers in modified theories of gravity is extremely challenging in Numerical Relativity, it is possible to effectively model echoes within linear perturbation theory (e.g., \cite{Burgess:2018pmm}). This provides a unique phenomenological window to search for deviations from GR, in spite of the uncertainty in the echo templates. 
As GW190521 holds the record for the most massive binary BH (BBH) merger and the loudest ringdown ever detected in GWs \cite{Abbott:2020tfl,LIGOScientific:2020ufj,Capano:2021etf} it can be mostly modelled perturbatively. These features make GW190521 the ideal candidate to-date to look for potential evidence for GW echoes \cite{LongoMicchi:2020cwm}, which may subsequently test many proposals for quantum nature of BHs (e.g., \cite{Cardoso:2016rao,Oshita:2019sat,Cardoso:2019apo,Hayden:2020vyo,Ikeda:2021uvc}).

Even though no evidence for deviation from GR has yet been found in LIGO/Virgo main events \cite{TheLIGOScientific:2016src,Abbott:2017vtc,Abbott:2018lct}, tentative evidence for post-merger echoes has been reported by \cite{Abedi:2016hgu,Conklin:2017lwb,Abedi:2018npz,Salemi:2019uea,Holdom:2019bdv} (with JA and NA as co-authors on some), but they remain controversial, with results depending on event and methodology \cite{Westerweck:2017hus, Ashton:2016xff, Abedi:2017isz, Abedi:2018pst,Salemi:2019uea,Abedi:2020sgg,Abedi:2020ujo,Abedi:2020sgg}.
For example, in \cite{Uchikata:2019frs}, using their suggested template, they found no evidence for echoes (p-values of 0.98 and 0.92, in their Table II), while p-values of 5.5\% and 3.9\% (their Tables III \& IV, for O1 and O2 events respectively) are reported for the original template proposed by \cite{Abedi:2016hgu}.

Motivated by this, many groups have directed their efforts to search for echoes using a diverse set of tools and methodologies. So far, these searches have employed mainly two types of strategies: Template-based methods  \cite{Abedi:2016hgu,Westerweck:2017hus,Nielsen:2018lkf,Uchikata:2019frs,Lo:2018sep,Abbott:2020jks,Lo:2018sep,Abbott:2020jks,Wang:2020ayy,Westerweck:2021nue} and model-agnostic (or coherent) methods \cite{Abedi:2018npz,Conklin:2017lwb,Salemi:2019uea,Holdom:2019bdv,Ren:2021xbe}.  Using these approaches, different groups have drawn a range of conclusions about the existence of echoes, from positive \cite{Abedi:2016hgu, Conklin:2017lwb, Abedi:2018npz,Uchikata:2019frs, Holdom:2019bdv} (with p-values within $0.002\%$-$5\%$), to mixed \cite{Westerweck:2017hus,Nielsen:2018lkf,Salemi:2019uea}, and negative evidence \cite{Lo:2018sep,Uchikata:2019frs,Tsang:2019zra,Abbott:2020jks,Wang:2020ayy,Westerweck:2021nue,Ren:2021xbe}. Moreover, the echoes first identified around 1 second after the binary neutron star merger GW170817 \cite{Abedi:2018npz}, coincided with the time expected for formation of a BH, subsequently inferred from electromagnetic follow-up \cite{Gill:2019bvq}.

These apparent discrepancies between different search strategies are not surprising, as different methods may be more or less effective in identifying echoes, depending on the underlying theory. For example, the search for GW190521 echoes by the LIGO/Virgo collaboration \cite{Abbott:2020jks} (based on methodology developed in \cite{Lo:2018sep}) extends to a region of parameter space with nonphysical GW energy (thus diluting the evidence; see below and Appendix \ref{ADA_method})\footnote{While \cite{Abbott:2020jks} does not explicitly quote their priors, their search is based on \cite{Lo:2018sep}, which used $\Delta t_{\rm echo}< 0.5$ sec, while the echo for GW190521 is expected at $1.2 \pm 2$ sec. This highlights the importance of using physical models as a powerful tool to guide searches for exotic physics.}.

In this paper, we adopt an approach to look for GW echoes by making use of different GW data analysis tools, and physical echo waveform templates. The key underlying assumption is to replace the classical event horizon by a membrane with quantum dynamics. This leads to the formation of an ``echo chamber'', where GWs can be trapped between the quantum membrane and the classical angular momentum barrier, but periodically leaking out to infinity \cite{LongoMicchi:2020cwm,Abedi:2016hgu,Cardoso:2016rao,Cardoso:2016oxy,Wang:2018gin}. If the quantum membrane sits at a proper distance of $l_{\rm QG}$ away from the would-be horizon, individual echoes will be temporally separated by: 

\begin{figure*}
\begin{minipage}[l]{0.43\textwidth}
\begin{subfigure}{1.0\textwidth}
  \centering
  \includegraphics[width=1.0\linewidth]{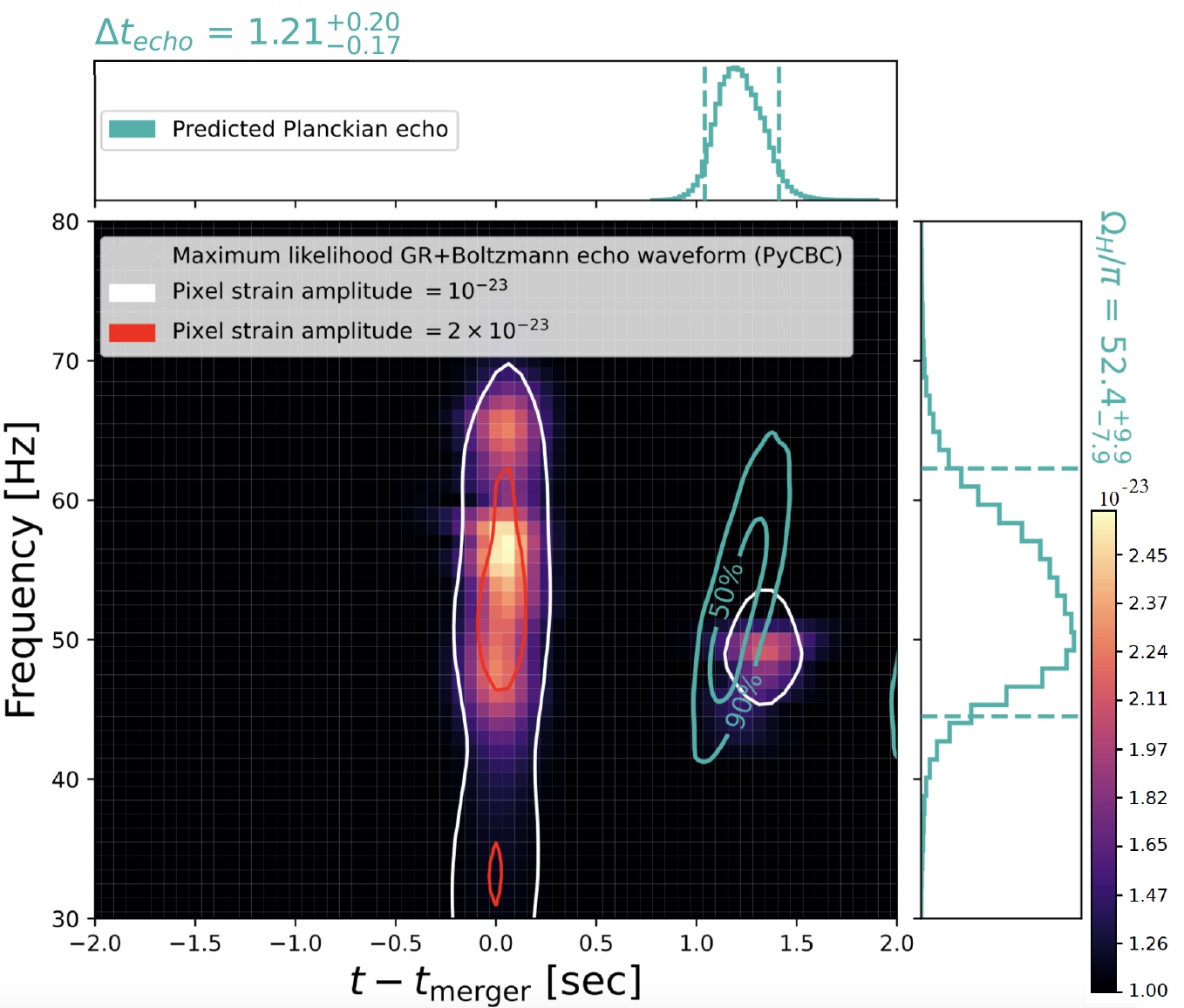}\hfill
  \caption{}
  \label{fig:sub-1}
\end{subfigure}
\begin{subfigure}{1.0\textwidth}
  \centering
  \includegraphics[width=1.0\linewidth]{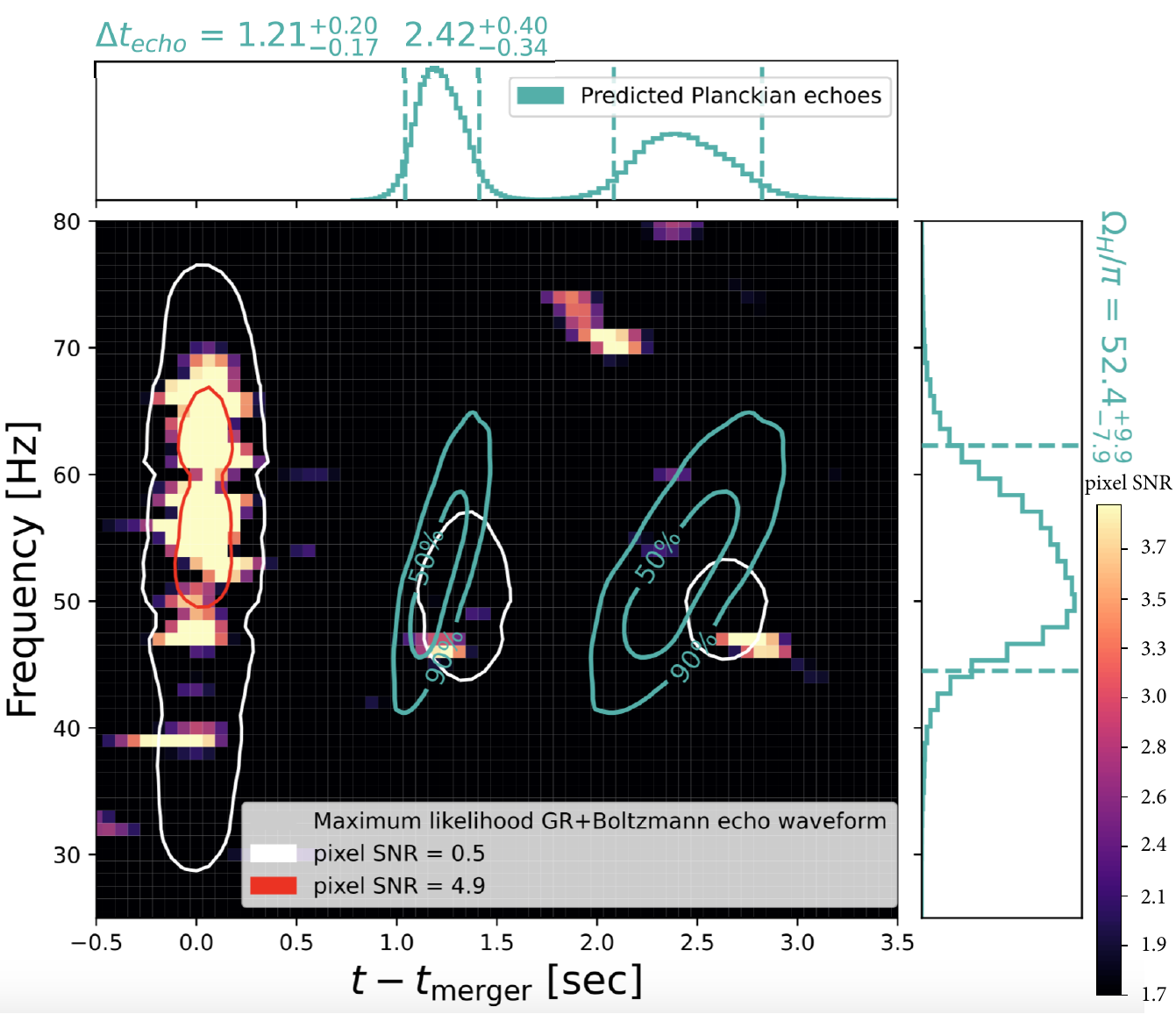}\hfill
  \caption{}
  \label{fig:sub-2}
\end{subfigure}
\begin{subfigure}{1.0\textwidth}
  \centering
  \includegraphics[width=0.85\linewidth]{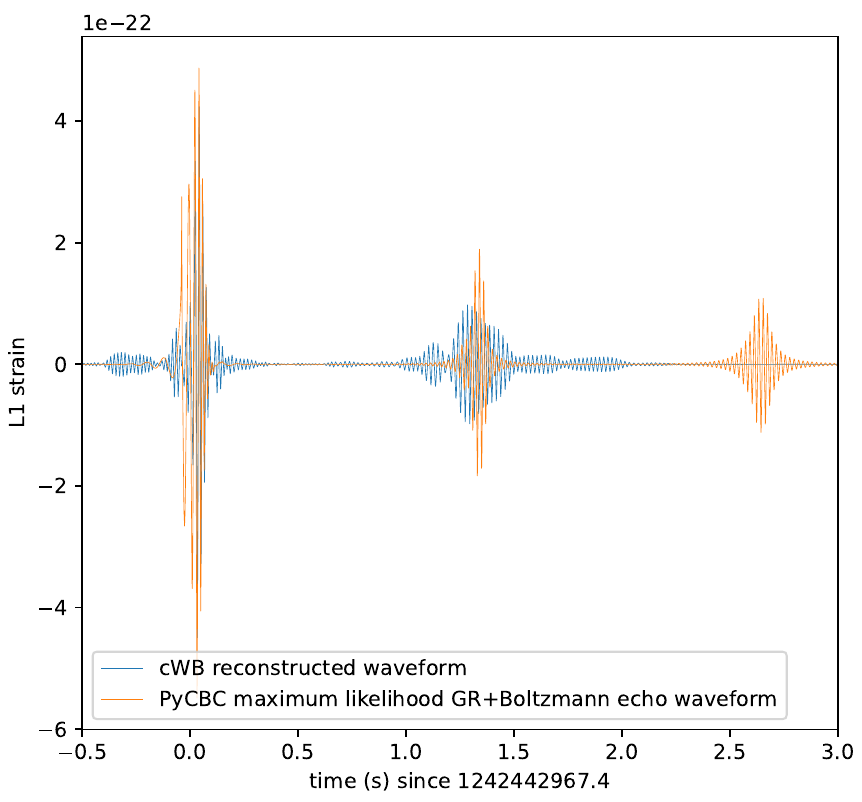}\hfill
  \caption{}
  \label{fig:sub-3}
\end{subfigure}
\end{minipage}
\hfill{}
\begin{minipage}[r]{0.4\textwidth}
\begin{subfigure}{1.0\textwidth}
  \centering
  \includegraphics[width=1.0\linewidth]{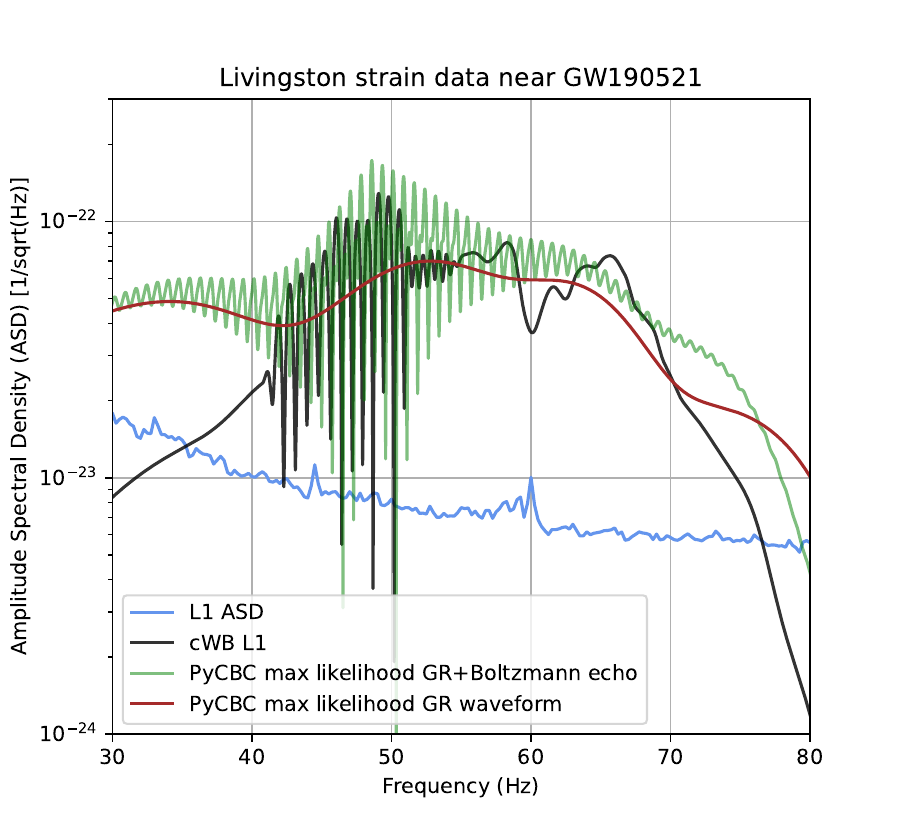}\hfill
  \caption{}
  \label{fig:sub-4}
\end{subfigure}
\begin{subfigure}{1.0\textwidth}
  \centering
  \includegraphics[width=1.0\linewidth]{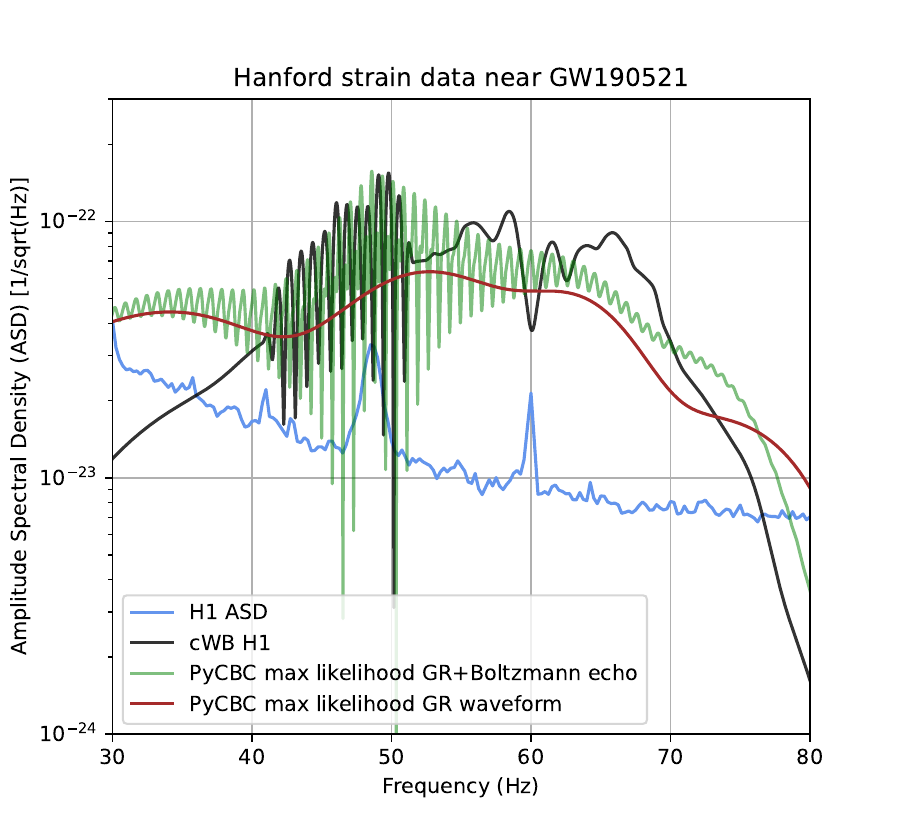}\hfill
  \caption{}
  \label{fig:sub-5}
\end{subfigure}
\begin{subfigure}{1.0\textwidth}
  \centering
  \includegraphics[width=1.0\linewidth]{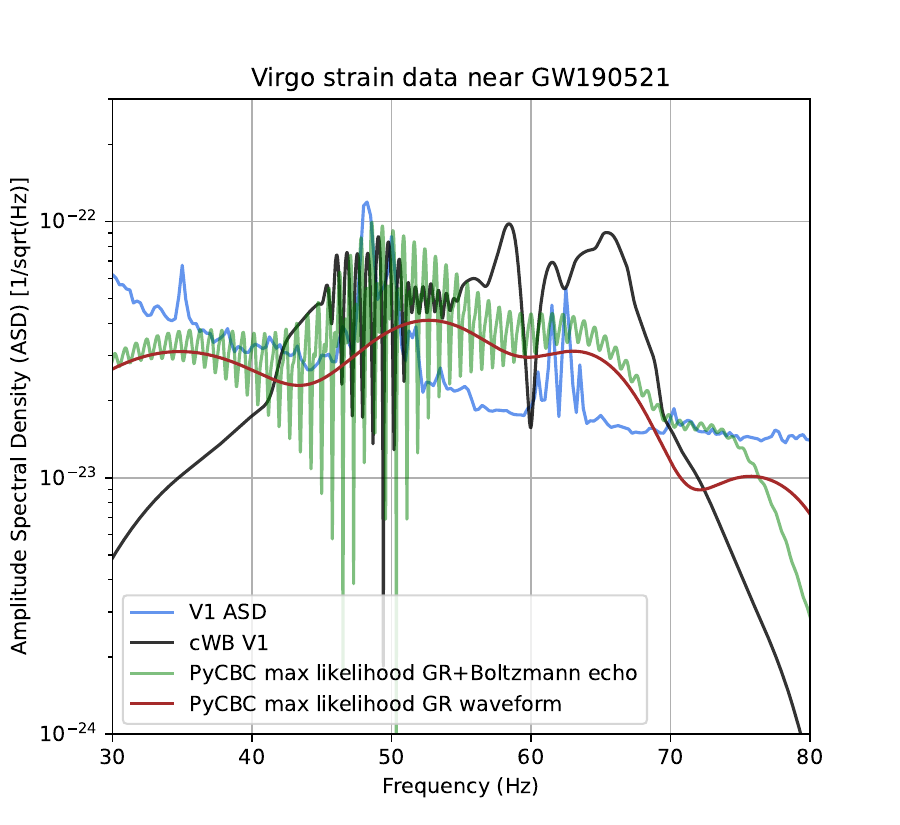}\hfill
  \caption{}
  \label{fig:sub-6}
\end{subfigure}
\end{minipage}
\caption{(a,b): White and red contours represent the maximum likelihood GR+Boltzmann echoes (NRSur7dq4). The light-sea-green histograms show the {\it predicted} 1D distributions for the 1st and 2nd Planckian echoes, based on the GR best-fit to GW190521. (a): cWB's reconstructed waveform spectrogram compared to the maximum likelihood GR+Boltzmann echo waveform (contours) using PyCBC assuming  Planckian echoes for stimulated Hawking radiation. (b): Whitened H1$\times$L1 cross-spectrogram and predicted Planckian echoes for stimulated Hawking radiation (see Appendix for details of this method). (c): Comparison between cWB's and PyCBC's waveforms. (d,e,f): Amplitude Spectral Density (ASD) from the reconstructed waveform as obtained by cWB (PyCBC). Here, the repeating pattern for echoes in real time leads to the oscillatory pattern in the frequency domain.}
\label{Fig_1}
\end{figure*}

\begin{figure*}[ht!]
\centering
\includegraphics[width=1.\textwidth]{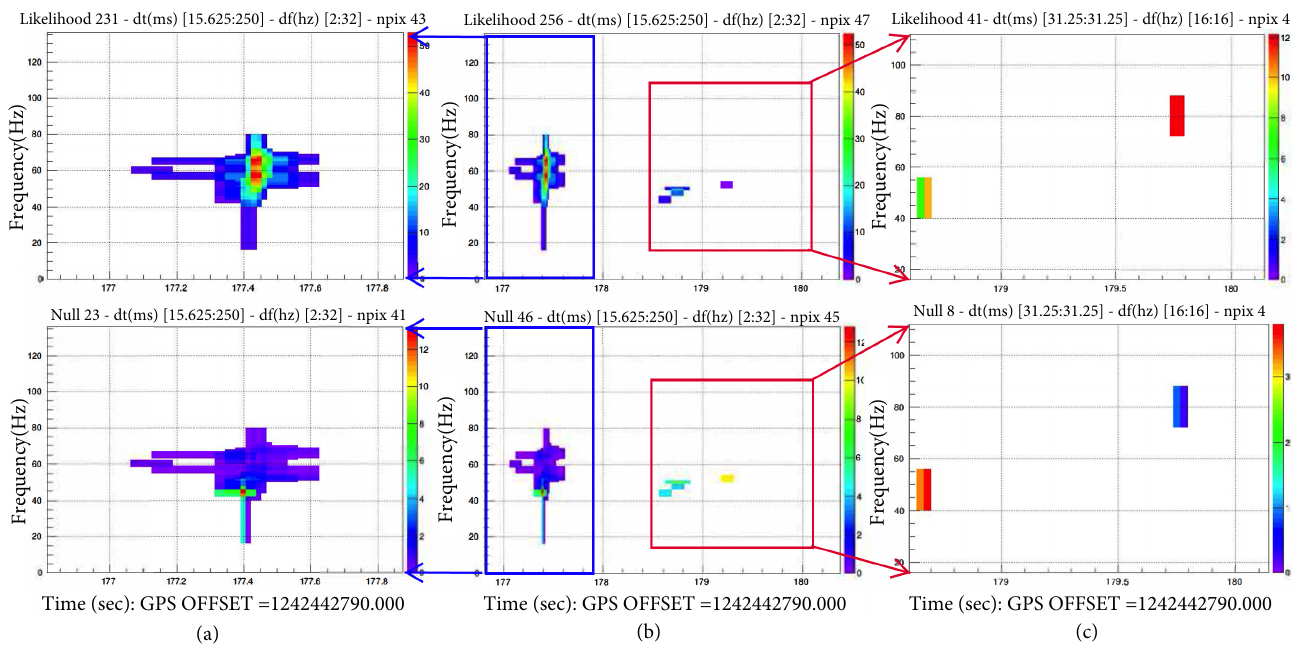}
\caption{cWB color-coded time-frequency maps for GW190521. Caption \textbf{(a)} (compare to \cite{GW190521cwb}) and caption \textbf{(b)} were obtained by using the recommended LVK thresholds on the original data. Caption \textbf{(c)} were obtained by using the weaker thresholds on the GR subtracted data. The middle panels differ from the left ones mainly by the time length of the search. Colorbars on top panels show the likelihood of individual pixels, which is related to the coherent energy between detectors, $E_C$, while the colorbars on bottom lines show the null energy $E_N$ (see text for details). }
\label{CWBpixels}
\end{figure*}
\begin{eqnarray}
\Delta t_{\rm echo} \simeq \frac{4 G M_{\rm BH}}{c^3}\left(1+\frac{1}{\sqrt{1-a^2}}\right) \times \ln\left(M_{\rm BH} \over \Lambda^{-1}M_{\rm planck}\right)\nonumber\\ \label{delay}
\end{eqnarray}
where $M_{\rm BH}$ and $a$ are the final redshifted mass and dimensionless spin of the BH remnant, while $\Lambda=l_{\rm{P}}/l_{\rm{QG}}$ represents the ratio of the energy scale of the quantum membrane physics to the Planck scale. While we expect $\Lambda\sim1$ for Planckian echoes, we do vary this parameter within 
\begin{equation}
-13\leq\log_{10} \Lambda\leq13 \label{priorlambda} \footnote{This may alternatively be attributed to $\pm10~\hbar k^{-1}T^{-1}_{H}$ uncertainty in stimulated emission time, where $T_H$ is the Hawking temperature.}
\end{equation}
For the  GW190521 BH mass and spin, we find $\Delta t_{\rm echo}\left|_{\Lambda=1}\right.=1.2\pm0.2$ sec (at 90\% confidence), which may explain why no echoes were found by the LIGO/Virgo collaboration \cite{Abbott:2020jks} assuming a  0.05 sec $\leq \Delta t_{\rm echo}  \leq$ 0.5 sec prior. We shall instead use the physical prediction (\ref{delay}), which fixes $\Delta t_{\rm echo}$ in terms of the final BH parameters and the scale of new physics. 

The next important ingredient is a physical model for the emission of echoes. If we consider Hawking radiation as the quantum {\it spontaneous} emission of light particles with $\hbar \omega \sim k T_H$ from a quantum BH (e.g., fuzzballs \cite{Lunin:2001jy,Lunin:2002qf}) in vacuum, it is natural to expect {\it stimulated emission} to happen, if we immerse the BH in a classical radiation field with similar frequencies. Indeed, \cite{Oshita:2019sat,Wang:2019rcf} proposed this as a natural mechanism to produce echoes of GWs, showing that quantum horizons should have a flux reflectivity, given by a Boltzmann factor, $\exp(-\hbar|\omega|/kT_H)$ (in their comoving frame), through quantum stimulated emission. For observers at infinity, we have to shift the frequency $\omega\to\omega-m\Omega_H$, where $\Omega_H$ and $m$ are the horizon angular frequency and azimuthal harmonic number, respectively. 

While one cannot produce a precise template without a quantum simulation of BBH merger, a plausible ansatz for the expected radiation is given as a sum over Boltzmann echoes:
\begin{eqnarray}
    h_{\rm GR+echoes}(\omega) =  h_{\rm GR}(\omega) \left[ 1+ A e^{i\phi} \sum_{n=1}^\infty {\cal R}^n \right], \label{eq:Boltzmann_template} \\ {\cal R} \equiv \mp \exp[-\frac{\hbar|\omega -2 \Omega_{H}|}{2kT_{H}} + i\omega \Delta t_{\rm echo} ], \label{eq:R}
\end{eqnarray}
where $A e^{i\phi}$ quantifies their overall amplitude, while the modulus and phase of ${\cal R}$ quantify their relative damping and temporal separation, respectively. 
Generally, we expect
\begin{equation}
0\lesssim A\lesssim2 \label{priorA}
\end{equation}
and
\begin{equation}
0< \phi\leq2\pi \label{priorphi}
\end{equation}
due to GR non-linearities. Furthermore, as is often done, we have assumed that the energy in BBH ringdown and echoes are dominated by quadrupolar radiation at $\ell=m=2$.

In this work, for the first time we search for a physical model of stimulated Hawking radiation, and use two well-established analysis packages in order to cross-validate our findings: the template-based PyCBC \cite{Biwer:2018osg,alex_nitz_2021_5347736} and model-agnostic coherent WaveBurst (cWB) \cite{klimenko_sergey_2021_4419902}. We thus can evaluate the statistical significance and energy of the post-merger signal in independent and complementary ways, also quantifying the evidence for the event and its post-merger trigger to be co-located. Data for these independent (including priors for all GR waveform parameters) searches available at \cite{GW190521echo}.

\section{Search for Stimulated Hawking Radiation}
We first used the PyCBC inference \cite{Biwer:2018osg} pipeline with the dynamic nested sampling algorithm {\it dynesty} \cite{Speagle_2020}.
The samplers serve as a way to map the hypothesis' posterior probability distribution in the parameter space for a given dataset. Provided the likelihood function and the prior, it is possible to measure the evidence in favor of (or against) the existence of a signal in the data. During this process, it is assumed a Gaussian background, and the three detector network H1-L1-V1 is used.

In order to quantify evidence for echoes in data, we first obtained the Bayes factor for a given template (based on GR, with and without echoes) against Gaussian noise. The Bayesian evidence for stimulated Hawking radiation, or Boltzmann echoes, is then given by the ratio of Bayes factors for $h_{\rm GR+echoes}(\omega)$ (Eq.\ref{eq:Boltzmann_template}) over the Bayes factor of the pure GR template, $h_{\rm GR}(\omega)$.  For the latter, we adopt the NRSur7dq4 surrogate waveform \cite{VarmaSurrogate}.

\begin{figure*}
\centering
\includegraphics[width=\textwidth]{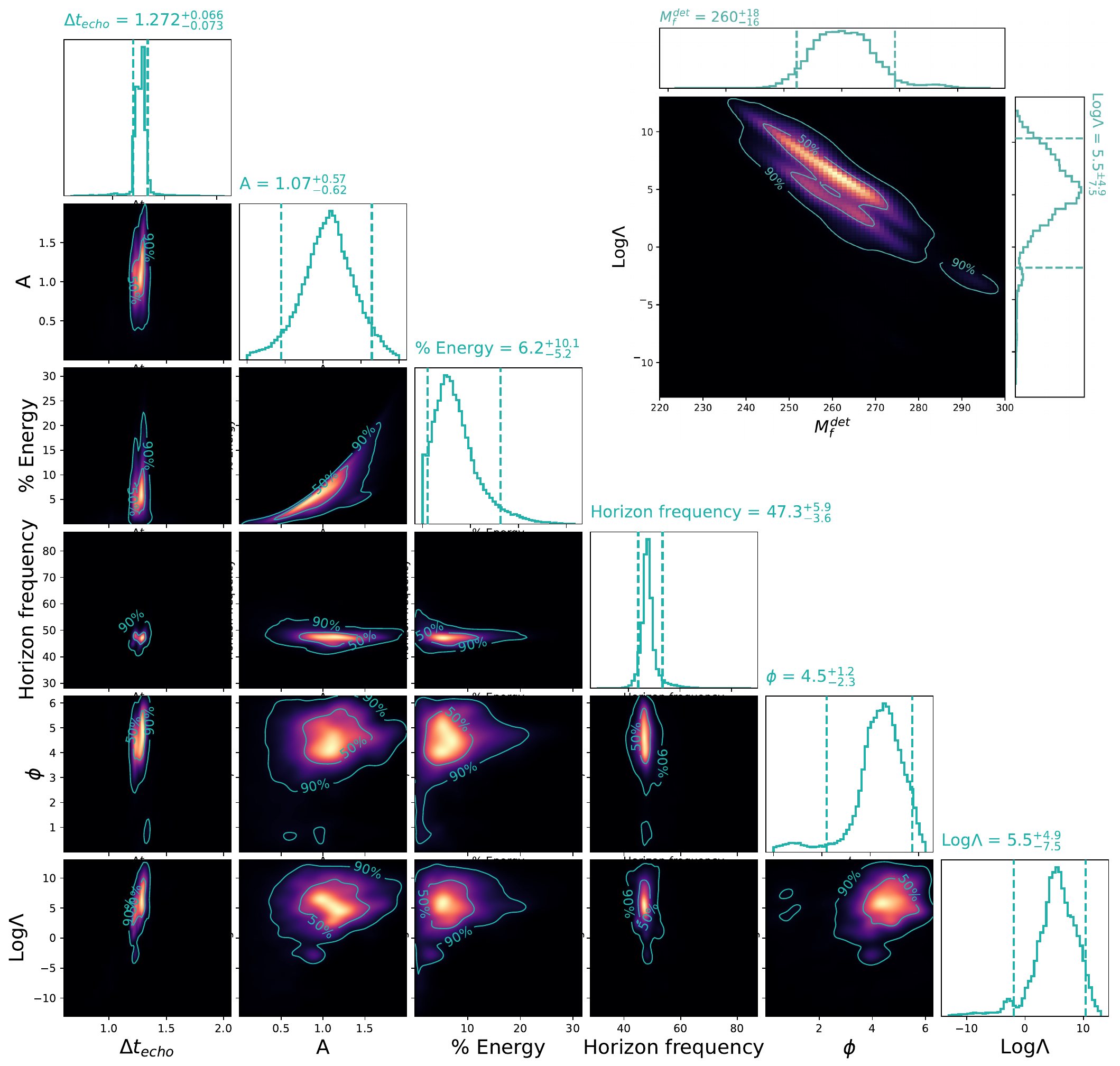}
\caption{Parameter estimation of Boltzmann echo waveform using NRSur7dq4. We find Bayes factor of $\mathcal{B}=7.5^{+3.8}_{-2.3}$ in preference for echoes.}
\label{posterior_1}
\end{figure*}

Our analysis suggests a statistical preference for stimulated Hawking radiation: Starting with different dynesty seeds, we obtain a Bayes factor of $\mathcal{B}= 7.5^{+3.8}_{-2.3}$ (Fig. \ref{posterior_1}), at 90\% of confidence. Results obtained by using different GR surrogate waveforms and/or echo models are reported in Appendix \ref{appendixC}. 
 We also noticed that, depending on the choice of GR waveform and/or priors, we may have preference for either Planckian (Fig. \ref{Fig_1}a) or super-Planckian echoes: $\log_{10}\Lambda=5.5^{+4.9}_{-7.5}$ or additional time delay of $(4.0^{+3.6}_{-5.5})~\hbar k^{-1} T^{-1}_H$  (Fig. \ref{posterior_1}; see also Appendix). This is due to degeneracy between $\Lambda$ and the final BH mass, which can be seen in Fig. \ref{posterior_1}.

\section{PyCBC false positive and true positive estimation}\label{PyCBCpvalue}
\begin{figure}
\centering
\includegraphics[width=0.5\textwidth]{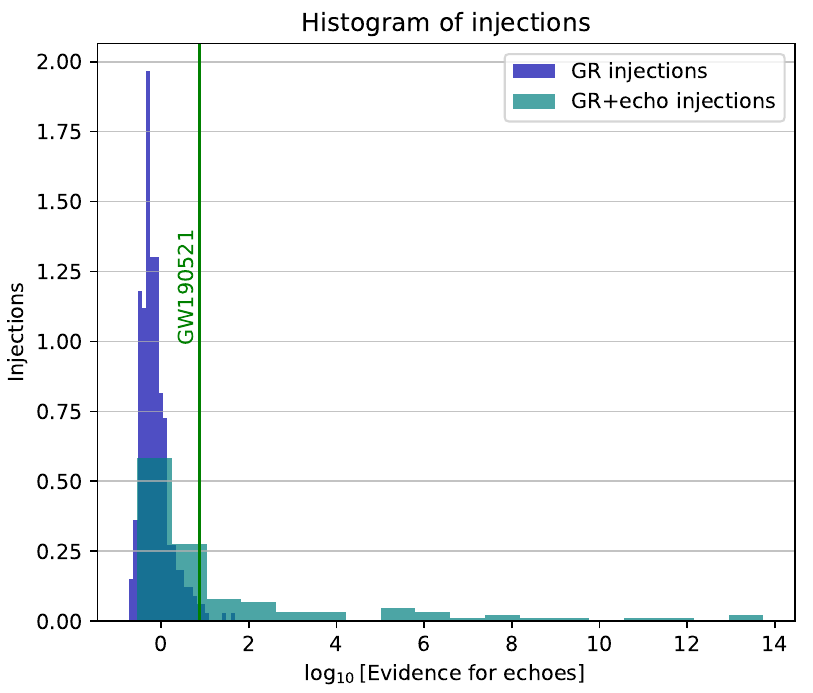}
\caption{Histograms to quantify PyCBC false positive (GR injections) and true positive (GR+echo injections). Comparing to the GW190521 echo, we obtain their values as $1.46^{+1.17}_{-0.875} \%$, and $34.5\pm 7.3\%$ respectively.\label{hist_pycbc}}
\end{figure}

We quantify the false and true positive detection probabilities using random samples drawn from PyCBC GR posterior, obtained from 10 different seed runs. To assess the true positive we have to inject models containing echoes. In this case, we construct these waveforms by adding two echoes to the previously selected GR templates. The additional echoes are randomly constructed from uniform priors of (\ref{priorlambda}), (\ref{priorA}), and (\ref{priorphi}). They are then injected at random times within $-32 \rm{\ sec}<t-t_{merger}<32 \rm{\ sec}$, excluding $-5 \rm{\ sec}<t-t_{merger}<5 \rm{\ sec}$ (due to presence of event). To compare the foreground (true positive) to the background (false positive), we also inject a sample with zero echo amplitude. Then we run the same script for injections with GR+echo and GR waveform (which is same as what we did for GW190521). The number of injections performed for zero amplitude echoes (GR waveforms) is 343, while this number is 110 for non-zero amplitude echoes (GR+echo waveform). Fig. \ref{hist_pycbc} shows the results of this analysis in a histogram plot. We find that $1.46^{+1.17}_{-0.88} \%$($34.5\pm 7.3\%$) of GR only (GR+echo) injections yield larger echo Bayes factors than 7.5, which was the value we obtained for GW190521. The ratio of these two numbers yields an empirical likelihood $= 24.3^{+42.7}_{-11.8}$, for the GR+echo model compared to the GR-only hypothesis. 


\section{Coherent WaveBurst (cWB) search}
As  an independent method, we then used the model-agnostic cWB package to test the robustness of the evidence for echoes. The cWB package is a pipeline specialized in reconstructing GW signals by making minimal assumptions regarding the waveform morphology. To perform such reconstruction, we start by searching for burst of energy excess that are coherent (coherent energy $E_{C}$), in the time-frequency domain, in multiple detectors. During its workflow, the software  defines a coherent statistics that is related to the signal-to-noise ratio (SNR) in the detector. Once the waveform is estimated, it is subtracted from the data and the null energy $E_{N}$ is estimated. As a measure of the coherence of the signal the network correlation coefficient as $cc=E_{C}/(E_{C}+E_{N})$, which satisfies $0\leq cc\leq1$. The cc parameter is one of the multiple search thresholds that triggers have to satisfy, usually $cc>0.7$ is required to claim a potential detection. As final products, cWB is capable of producing a reconstructed  waveform and sky-location analysis.

We first reproduce GW190521 main event's detection (see Fig. \ref{CWBpixels}a) in agreement with Fig. 2 of \cite{GW190521cwb}. We reconstruct the spectrogram of the event+post-merger echoes using all the recommended search parameters \cite{cWB}, but allowing for separations of $\gtrsim$1 second (Fig. \ref{CWBpixels}b). In order to quantify their significance and sky-location, we can further subtract the best-fit GR template from the strain data and look for the echo-only cWB signal with lower thresholds (Fig. \ref{CWBpixels}c). The comparison between the reconstructed model-agnostic waveform (cWB) and the best-fit stimulated Hawking radiation waveform (PyCBC) can be seen in Figs. \ref{fig:sub-1} and \ref{fig:sub-2} (high thresholds) and  \ref{fig:sub-3} (low thresholds). We notice that (apart from a rogue pixel at $\approx80$Hz for the lower threshold search), both methods achieve qualitatively consistent post-merger waveforms. Given this analysis, we obtained not only the event' spectrogram but also its sky localization and its statistical significance, both the main event and post-merger signal (see Table \ref{Table_cwb}).

\begin{figure}
    \includegraphics[width=0.45\textwidth]{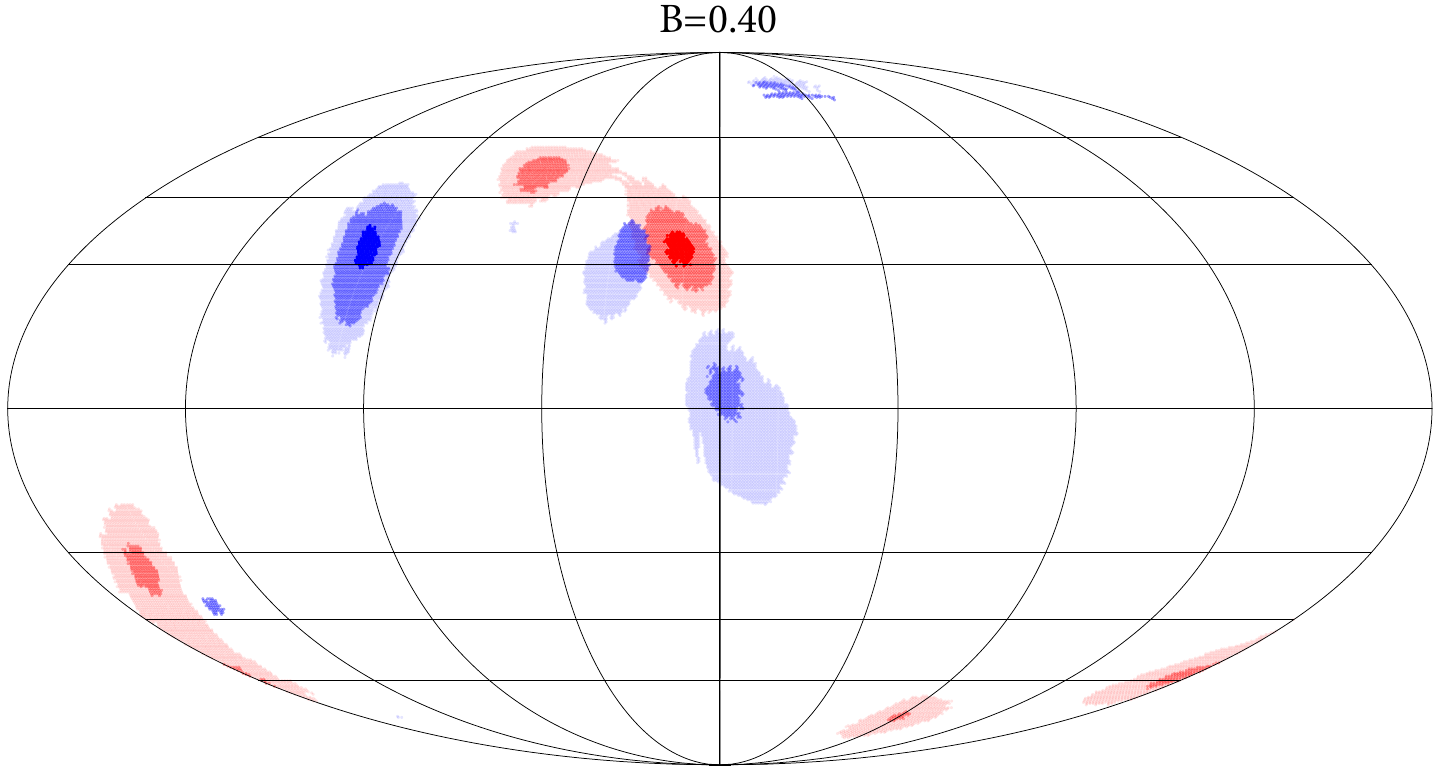}
    \includegraphics[width=0.45\textwidth]{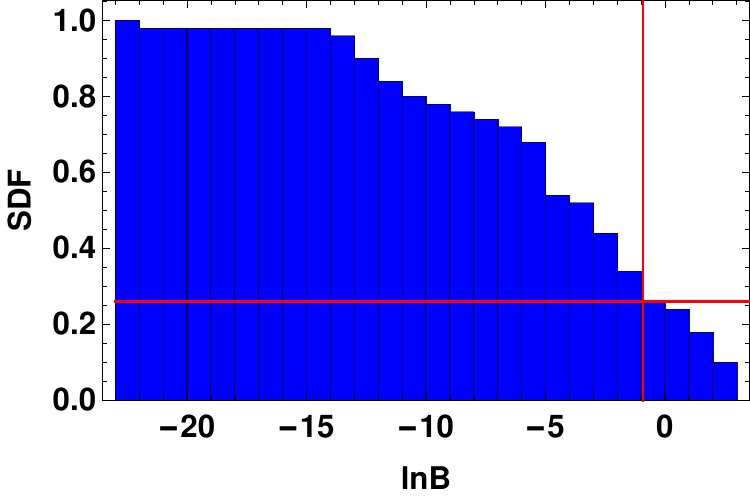}
    \caption{Top: Sky localization comparison. Here the different opacities represent the 10-50-90\% confidence regions in the sky. Red and blue areas represent the main event and echo (obtained after subtracting the main event), respectively. We notice that our main event results are in agreement with \cite{GW190521cwb}. Sky localization analysis for cWB echoes, showing inconclusive evidence for/against the co-localization of the main event and the echoes, with a Bayes factor of ${\cal B}^{\rm co-loc} \simeq 0.4$. Bottom: Cumulative distribution function for the co-localization Bayes factors for all noise triggers found during the p-value estimation (see Appendix B). Red lines mark the Bayes factor found for the post-merger signal. The PDF holds a mean of $\approx-4.87$ and a standard deviation of $\approx 5.69$.\label{skyloc}}
\end{figure}

\begin{table}[]
    \centering
\begin{tabular}{|c||c|c|c|c|c|c||c|}
        \hline
         & \multicolumn{3}{c|}{cWB $\mathcal{B}^{\rm{co-loc}}_{ab}$} &\multicolumn{3}{c||}{cWB} & PyCBC \\
         \hline
         & Joint & Main event & Echo & $\rho$  & cc  & $\rm SNR$ & $\rm{SNR}$ \\
        \hline
        \hline
         Joint  & 51.88  & 46.20 &  0.32 & 10.6 & 0.90 &  16.0 & 16 \\ 
        \hline
        Main event & 46.20 & 42.51 & 0.40 & 10.8  & 0.95 & 15.2 & 15 \\ 
        \hline
        Echo & 0.32 & 0.40 & 37.98 &  4.1 & 0.88 & 4.1 & 6\footnote{Based on search for echoes without including main event} \\ 
        \hline
    \end{tabular}
    \caption{Collection of several statistical findings. Here $cc$ quantifies the degrees of signal coincidence amongst detectors ($0\leq cc\leq1$), while $\rho$ is the so-called the effective correlated SNR (see  \cite{Salemi:2019uea}). While the evidence of 0.4 for co-localization is inconclusive, we notice an improvement of $\approx22\%$ in the Bayes factor for self-localization of the event when including the post-merger signal.}
    \label{Table_cwb}
\end{table}

While the signal-to-noise ratio for echoes is not very large (4.1/6 for echoes in cWB/PyCBC, compared to 15 for the main event; see Table \ref{Table_cwb}), it is in line with what is expected for stimulated Hawking radiation from an event with SNR$_{\rm ringdown}\sim16$, which is SNR$_{\rm echo}$=2-5(Fig. 9 in\cite{LongoMicchi:2020cwm} for mass ratio $q\sim0.8$\cite{Abbott:2020tfl}) 

We also estimate the background and obtain p-value of $5.1\times10^{-3}(2.6\sigma)$, (see Appendix \ref{SMB} for details) via an injection study.

\section{Energy estimation of echoes in cWB}\label{EnergyError}

One of cWB's capabilities is the reconstruction of the signal based on the excess of coherent energy in the network detectors. During our analysis of GW190521 we perform such reconstruction, which is shown in Fig. \ref{Fig_1}c. In this process, cWB reconstructs the waveform as seen by each one of the network detectors. In order to estimate the energy of the physical strain, we perform the following calculation:
\begin{equation}\label{energyestimate}
    E \approx\bigintss f^{2}\dfrac{\sum_{i}  |h_{i}(f)|^{2}S_{i}^{-2}(f) df }{\sum_{i}S_{i}(f)^{-2}df}
\end{equation}
where the sum is over the detector network, $S_{i}(f)$ and $h_{i}(f)$ is the noise ASD and the observed strain at the $i$-th detector respectively. The integral was performed between 0 and 100 Hz, which is a reasonable range, given the results found in Fig. \ref{fig:sub-second}. Equation (\ref{energyestimate}) is a way of weighting (or whitening) the contribution of the individual detector regarding its strain sensitivity. In order to compute the percentage of energy in the echo part of the waveform we compute:

\begin{equation}\label{Eper100}
    E_{echo}(\%) =\dfrac{E_{full}-E_{main}}{E_{main}},
\end{equation}
which is the value quoted in Table \ref{EnergyCWB} as equal to $13.8\%$. As this value was reconstructed via cWB we will refer to it as recovered energy. The next task is to estimate the error on the ``true'' energy, given the value of the recovered energy.

Our starting point for estimating the error on $E_{echo}(\%)$ is injecting 600 waveforms in noisy data around GW190521 event ($\pm32$ s.), after subtracting the main event. The injected waveforms $h_{i}^{inj.}(t)$ are constructed by using the reconstructed strain by cWB $ h_{i}^{rec.}(t)$ as follows:

\begin{equation}\label{injections}
    h_{i}^{inj.}(t) = h_{i}^{rec.}(t)(1 + \alpha \Theta(t-t_{1})\Theta(t_{2}-t))
\end{equation}
for $t_{2}>t_{1}$. In (\ref{injections}), the i index represents the detector, $\Theta(t)$ is the Heaviside step function and $t_{2}=t_{echo}+1.25$ and $t_{1}=t_{echo}-0.75$ are chosen such that only the echo part of the waveform is modified. The extra parameter $\alpha$ is sampled from -1 to 2, in a way to be more densely populated between -1 and 0.5. As the energy is quadratic on the strain, our method is equivalent to inject wave with energies of:

\begin{equation}\label{Einjections}
    E_{echo}^{inj.}(\%) =  E_{echo}^{rec.}(\%)(1 + \alpha)^{2},
\end{equation}
where $E_{echo}^{rec.}(\%)=13.8\%$, as obtained from cWB in real data.

\begin{figure*}
    \centering

    \includegraphics[width=0.32\textwidth]{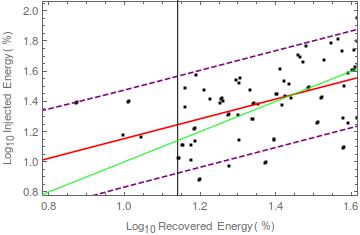}
    \includegraphics[width=0.32\textwidth]{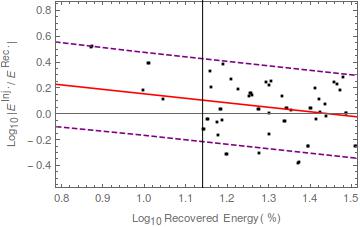}
    \includegraphics[width=0.32\textwidth]{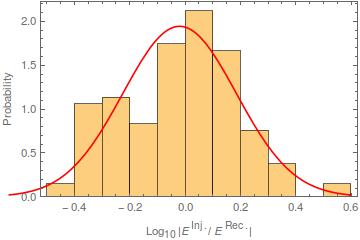}
    
    \caption{Left: Energy Plot (log scale). Injected energy as a function of recovered energy for the injection study. The green line is a line of slope 1 for inserted for comparison reasons. Right: Residual Plot (log scale). Difference between recovered and injected energy as a function of  recovered energy. Both: Solid red and dashed purple lines represent the best fit of the data to a linear function and its 90\% confidence region, respectively. The vertical solid black line marks recovered energy of $13.8\%$, the value encountered for GW190521. Right: Histogram distribution of the logarithm of the ratio of injected to recovered energy. Our error estimations based on the Gaussian approximation of the histogram lead us to a measurement of $13^{+16}_{-7}$\% of recovered energy. }
    \label{fig:Energy_error}
\end{figure*}

After performing the injections, we perform similar searches to the one leading to Fig. \ref{CWBpixels}b and computed the echo energy for each injection via (\ref{energyestimate}) and (\ref{Eper100}). Our goal is to determine the dispersion of the measured energy, provided a prior knowledge of the existing signal on the data. After performing the search with  selecting only events which reconstructed echo energy above between 5 and 63\%. This selection was made in order to eliminate the case with no clear echo detection and to avoid selection bias due to our abrupt cut on the injected energies. This way we selected 132 out of the 600 original injections. Fig. \ref{fig:Energy_error} shows our main results.

From the left and middle panel of Fig. \ref{fig:Energy_error}, we notice that weak(strong) signals are often under(over)-estimated. From the right caption of the same figure we can see that the dispersion of the measurement can be well approximated by a Gaussian. From this Gaussian approximation we can estimate the 90\% confidence region of the dispersion histogram. The reason for using a log scale to estimate the confidence region is to avoid negative energy values. Using this method we estimate our energy estimation to be  $13^{+16}_{-7}\%$, in agreement with $6.2^{+10.1}_{-5.2}\%$ from PyCBC.

\begin{table}[]
    \centering
    \begin{tabular}{|c|c|c|c|c|}
         \hline
        $E^{\rm{Echoes}}_{\rm{Main Event}}$ & L1 & H1 & V1 & Mean \\ \hline \hline
         cWB$^{Rec.}$ & 15.3$\%$ & 10.8$\%$ & 6.6$\%$ & $13.8\%$  \\
         \hline
          cWB$^{Inj.}$ & - & - & - &$13^{+16}_{-7}\%$ \\ \hline 
          PyCBC & - & - & - & $6^{+10}_{-5}\%$ \\
          \hline
    \end{tabular}
    \caption{Energy ratio of the first echo with respect to the main event. cWB estimate was obtained from the reconstructed frequency-domain waveforms as in Fig. \ref{CWBpixels}a and \ref{CWBpixels}b, and the mean of the energies is given by weighting the waveforms by the noise ASD$^2$, see (\ref{energyestimate}).}
    \label{EnergyCWB}
\end{table}
\section{Sky co-localization}
Using the cWB sky localization posterior map, we could compute the Bayes factor for the co-localization hypothesis for signals $a$ and $b$ as:
\begin{equation}\label{eq:Bcoloc}
    \mathcal{B}^{\rm{co-loc}}_{ab} = N \sum_{i=1}^{N} P_{a}(\phi_{i},\theta_{i})P_{b}(\phi_{i},\theta_{i})
\end{equation}
where $P_{a}(\phi_{i},\theta_{i})$ is the posterior probability of event $a$ being located within the i-th of the N pixels of the skymap Appendix \ref{SMA}. We validated this method by injecting waveforms containing echoes. For loud echo injections, we find positive evidence for co-localization. However, as to the real data, we find an ``inconclusive'' Bayes factor of $0.4$ for the co-localization of cWB main event and echoes, see Table \ref{Table_cwb} and Fig. \ref{skyloc}. As we noted above in Fig. \ref{CWBpixels}c, we notice that lowering the cWB thresholds picks up a possible noise feature at 80 Hz (absent for the high threshold search), which could be interfering on our sky localization analysis. To avoid this, we can compare the Bayes factors for the self-localization (diagonal entries on Table \ref{Table_cwb} cWB $\mathcal{B}_{ab}^{co-loc}$) for the joint analysis (event+echo) and the main event. Addition of the echo improves self-localization Bayes factor by $\approx22\%$. This suggests that excluding the 80 Hz noise feature (as in Fig. \ref{CWBpixels}b) may lead to marginal evidence for co-localization hypothesis.

\begin{figure*}
    \centering
    \includegraphics[width=0.45\textwidth]{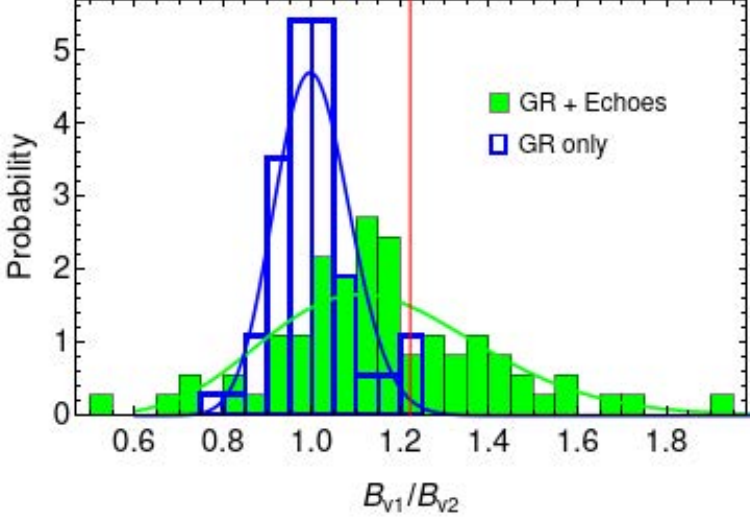}
    \includegraphics[width=0.45\textwidth]{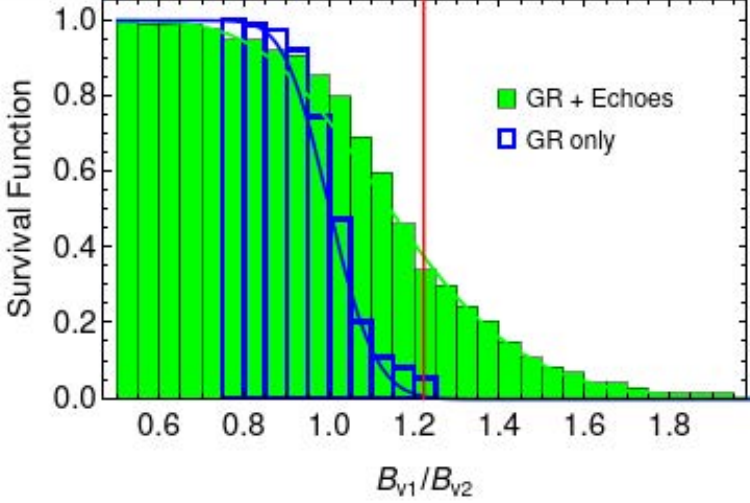}
    \caption{Injection study investigating the impact of the inclusion of pixels due to an echo detection. On the left we show the probability as a function of the co-localization Bayes factor ratio, whereas in the right we show the survival function. Here $B_{v1}(B_{v2})$ is the co-localization Bayes factor found using the same configuration files that lead to the data in Fig. \ref{CWBpixels}b(a)}
    \label{fig:coloc_inject}
\end{figure*}

In order to assess whether or not the 22\%  improvement in the self-localization Bayes factor is due to the inclusion of the echo-like pixels found in Fig. \ref{CWBpixels}b, we performed an extra analysis in Fig. \ref{fig:coloc_inject} (similar to Fig. (\ref{hist_pycbc}). In this second step, we performed 173 injections of echoes into the 32 seconds of LIGO/Virgo data stream, retrieved prior to the time of GW190521. We injected cWB's reconstructed waveform (as in Fig. \ref{CWBpixels}b) but with the echo amplitude multiplied by a random factor drawn from the same distribution described in Fig. \ref{fig:Energy_error}. For each one of the injections we repeated the procedures that lead us to Fig. \ref{CWBpixels}a and Fig. \ref{CWBpixels}b. From the 173 original injections only 74 yield a positive echo detection. Note that this number of positive echo detections supports a true positive factor of 43\% (which is comparable with true positive $=35\pm 7$\% obtained by PyCBC).  The self-localization was computed for these 74 events excluding and including echoes. The ratio of these two Bayes factors were computed, in a similar fashion to the one used to find the aforementioned 22\% factor. The distribution of these values can be found in Fig. \ref{fig:coloc_inject}.

In Fig. \ref{fig:coloc_inject}, the red line marks the value of 1.22, which represents the improvement found when performing the same analysis on real data. We estimated each probability density function as a Gamma distribution. After this estimation we then compute the odds ratio test to assess how likely each type of injection is to provide a 22\% improvement in the self-localization analysis. Comparing the green to the blue curve in Fig. \ref{fig:coloc_inject}, we find:

\begin{equation}\label{colocimprove}
    \left.\dfrac{P_{GR+echo}(\alpha)}{P_{GR}(\alpha) } \right|_{\alpha = 1.22} \approx 8^{+4}_{-3}.
\end{equation}

The error margins in (\ref{colocimprove}) were obtained by a bootstrap method and represent the 50\% confidence level. This result implies that the improvement of 22\% in self-localization found in the real data analysis is 8 times more likely to be found for data with injected echoes than for the case of picking up random noise (in GR-only injections). It should be noticed that, this likelihood ratio of 8 is not a direct measure of the co-localization and that our direct measures are still inconclusive.

Let us comment on why we consider the 80 Hz feature appearing in Fig. \ref{CWBpixels}c as possible noise, while considering the 50 Hz as a possible echo trigger. An attentive reader may also have noticed that there is another rogue pixel at 50 Hz around 179.2s in Fig. \ref{CWBpixels}b (missing from \ref{CWBpixels}c). In order to explain this seemingly arbitrary choice, we should first discuss the difference between  Figs. \ref{CWBpixels}b and \ref{CWBpixels}c. In Fig. \ref{CWBpixels}c, we not only lower the threshold of detectability but also change the assumption about how the signal should behave in the spectrogram space. While searching for the full signal we assume that the signal should chirp, meaning that the pixels should appear as an ascending diagonal in the spectrogram. This is a natural assumption to make when searching for coalescence signals, as they display a chirping morphology. One can notice that in Fig. \ref{CWBpixels}b the echo trigger also appears with a diagonal trend due to this assumption. While this is expected for the main GR signal, we do not know in principle the morphology of echoes in the spectrogram space. Hence we relaxed the chirp-like morphology assumption for Fig. \ref{CWBpixels}c (which had GR signal subtracted), where we searched for individual pixels surpassing the required thresholds. We then note that the two searches, with different thresholds and patterns, only have one trigger in common at $\approx$(178.5s,50Hz) (thus a robust echo trigger). In contrast, the other two triggers, [$\approx$(178.5s,50Hz) Fig. \ref{CWBpixels}b and $\approx$(179.8s,80Hz) Fig. \ref{CWBpixels}c] are not robust to these changes, and are considered ``rogue''. 

\section{Conclusion}

Using two independent and complementary approaches to GW search (PyCBC and cWB) we found positive evidence for the presence of stimulated Hawking radiation, or post-merger Boltzmann echoes in the aftermath of GW190521 BBH merger. Bayesian evidence for echoes is $7.5^{+3.8}_{-2.3}$ and false detection probability is $1.46^{+1.17}_{-0.875} \%$ where the ratio of the foreground probability to the background yields an empirical likelihood ratio of $24.3^{4+2.7}_{-11.8}$ (via PyCBC). An independent cWB implementation provides a frequentist p-value of $0.5\%$. However, at the current level of signal-to-noise ratio, the evidence for the co-localization of the echoes and the main event in the sky remains inconclusive.

We conclude by noting that no indisputable evidence ($\geq5\sigma$) for stimulated Hawking radiation from BH mergers, or Boltzmann echoes, were to be expected at current levels of GW detector sensitivity \cite{LongoMicchi:2020cwm}. From a conservative standpoint, current analysis provides the best constraint on the energy and morphology of this signal to date. The fact that the signal is robust to the use of independent methods, alongside our extensive tests of false and true positives using injections in LIGO/Virgo data, should lend credence to the presence of stimulated Hawking radiation in the aftermath of GW190521.  Therefore, we conclude that echoes should be a prime target for the next generation of GW detectors.

\begin{acknowledgements}
We thank Cecilia Chirenti, Randy Conklin, Cole Miller, Germano Nardini, Alex B. Nielsen,  and Francesco Salemi for useful discussions. JA thanks the Max Planck Gesellschaft and the Atlas cluster computing team at AEI Hannover for support and computational help. JA was supported by ROMFORSK grant Project. No. 302640. LFLM thanks the financial support of the S\~ao Paulo Research Foundation (FAPESP) grant 2017/24919-4 and of Coordenação de Aperfeiçoamento de Pessoal de Nível Superior- Brasil (Capes) - Finance code 001 through the Capes-PrInt program, and thanks Perimeter Institute for the access to its cluster. NA is supported by the University of Waterloo, Natural Sciences and Engineering Research Council of Canada (NSERC) and the Perimeter Institute for Theoretical Physics.
Research at Perimeter Institute is supported in part by the Government of Canada through
the Department of Innovation, Science and Economic Development Canada and by the
Province of Ontario through the Ministry of Colleges and Universities. This research has made use of data, software and/or web tools obtained from the GW Open Science Center (https://www.gw-openscience.org), a service of LIGO Laboratory, the LIGO Scientific Collaboration and the Virgo Collaboration. LIGO is funded by the U.S. National Science Foundation. Virgo is funded by the French Centre National de Recherche Scientifique (CNRS), the Italian Instituto Nazionale della Fisica Nucleare (INFN) and the Dutch Nikhef, with contributions by Polish and Hungarian institutes.
\end{acknowledgements}

\bibliography{arXiv}



\appendix 
\addcontentsline{toc}{section}{APPENDICES}




\section{Co-localization}\label{SMA}

Here, we prove (\ref{eq:Bcoloc}). The two hypotheses are: 
\begin{enumerate}
    \item $\rm H_{0}$: The two events are not co-localized
    \item $\rm H_{1}$: The two events are co-localized
\end{enumerate}

The Bayes factor is defined as: 
\begin{align}
    \mathcal{B}^{\rm{co-loc}} &\equiv \dfrac{{\cal L}(\rm data| H_{1})}{{\cal L}(\rm{data | H_{0}})}  \\&= \dfrac{\sum_{p_1}  {\cal L}({\rm data|H_{1}},p_1)P({\rm H_1}, p_{1})}{\sum_{p_0}  {\cal L}({\rm data|H_{0}},p_0)P({\rm H_0}, p_{0})},  \nonumber
\end{align}
where ${\cal L}$ functions quantify the likelihood of recovering ``data'' within a given hypothesis, while $p_0$ and $p_1$ are the parameters of the models, with priors $P(...)$, within hypotheses ${\rm H_0}$ and ${\rm H_1}$, respectively. In particular, $p_0$ would be the sky locations of the two independent events $(\phi_i,\theta_i,\phi_j,\theta_j)$, while $p_1$ is the sky location of a single event $(\phi_i,\theta_i)$. Therefore, the likelihood functions are related to the sky maps:
\begin{eqnarray}
{\cal L}({\rm data|H_{0}},p_0) = P_1(\phi_i,\theta_i) P_2(\phi_j,\theta_j), \\
{\cal L}({\rm data|H_{1}},p_1) = P_1(\phi_i,\theta_i) P_2(\phi_i,\theta_i).
\end{eqnarray}
Meanwhile, assuming uniform priors across the sky in either case implies: 
\begin{equation}
P({\rm H_0}, p_{0}) = N^{-2}, P({\rm H_1}, p_{1}) = N^{-1},
\end{equation}
for $N$ pixels in the sky. Therefore,
\begin{align}\label{BcolocDerive}
    \mathcal{B}^{\rm{co-loc}} &= \dfrac{N \sum_i P_{1}(\phi_{i},\theta_{i})P_{2}(\phi_{i},\theta_{i})}{  (\sum_i P_{1}(\phi_{i},\theta_{i}))\times (\sum_j P_{2}(\phi_{j},\theta_{j}))} \nonumber \\
    & = N \sum_{i=1}^{N} P_{1}(\phi_{i},\theta_{i})P_{2}(\phi_{i},\theta_{i}),
\end{align}
where in the last line we used the fact that $P_{1}$ and $P_{2}$ are normalized across the sky. \qed

We should also note that in our derivation of (\ref{BcolocDerive}) it was implicitly assumed that the pixels are of equal area. Nevertheless, this assumption is validated by the fact that cWB makes use of the HEALPix software \cite{healpix}, which performs a Hierarchical Equal Area isoLatitude Pixelation of the sky sphere.

The sky probability maps for GW190521 main event and echo can be compared in Fig. \ref{skyloc}a. In Fig. \ref{skyloc}b, we show the survivor density function found for the Bayes factors of the 51 triggers due to noise found during our p-values estimation (see Appendix \ref{AppB}).

\section{Co-localization GW151012 and GW151226\label{AppB}}

During our analysis we were also aware of the results obtained in \cite{Salemi:2019uea}, reporting searches for post-merger signals following GWTC-1 events using the cWB package. Among the analysed events, two of them deserve extra attention: GW151012 and GW151226. According to the authors of \cite{Salemi:2019uea}, these events showed post-merger signals with p-values of $.0037\pm0.0014$ and $0.025\pm 0.005$, respectively. However, they concluded that the post-merger signal in GW151012 (despite its low p-value) is unlikely to be originated from the same sky position as the main event. This conclusion was based on the multi-peak structure found for the time-delay distributions of the secondary pulses, but no Bayes factors were calculated to test the co-localization hypothesis. In this section, we address this quantitatively. 

\begin{figure*}
    \centering
    \includegraphics[width=0.35\textwidth]{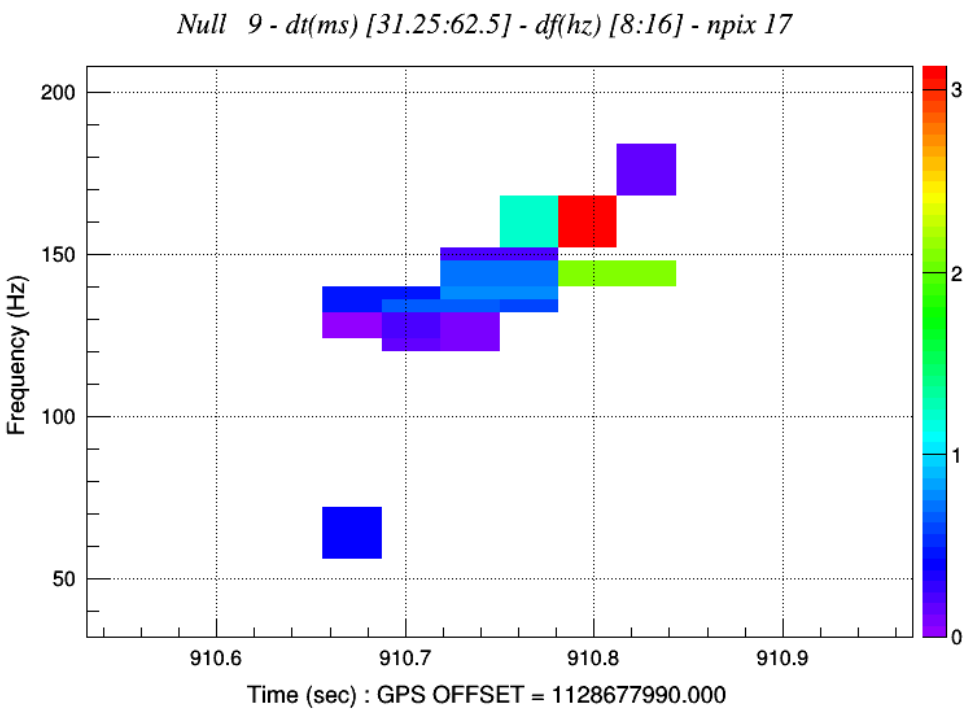}
    \includegraphics[width=0.35\textwidth]{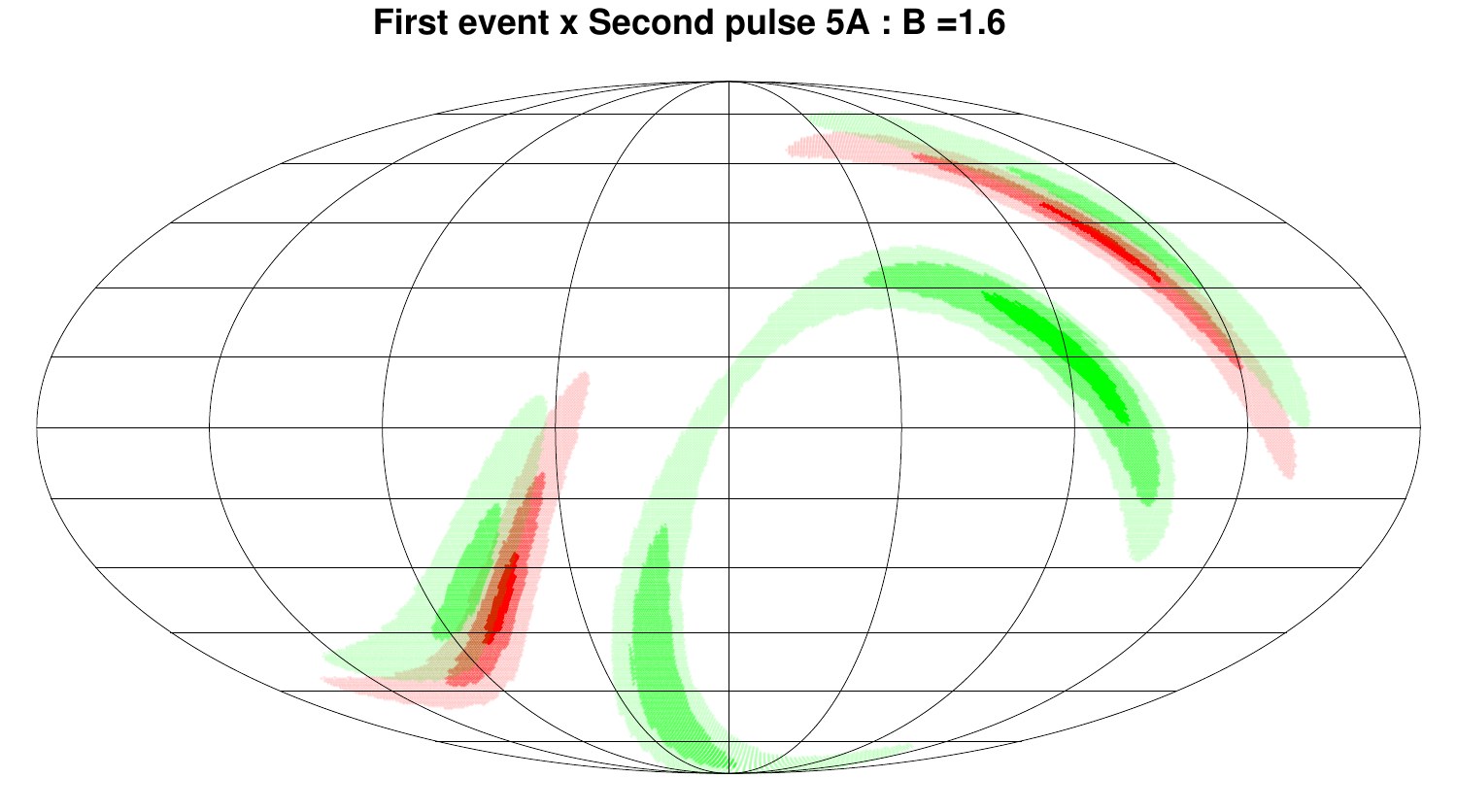}
    \includegraphics[width=0.35\textwidth]{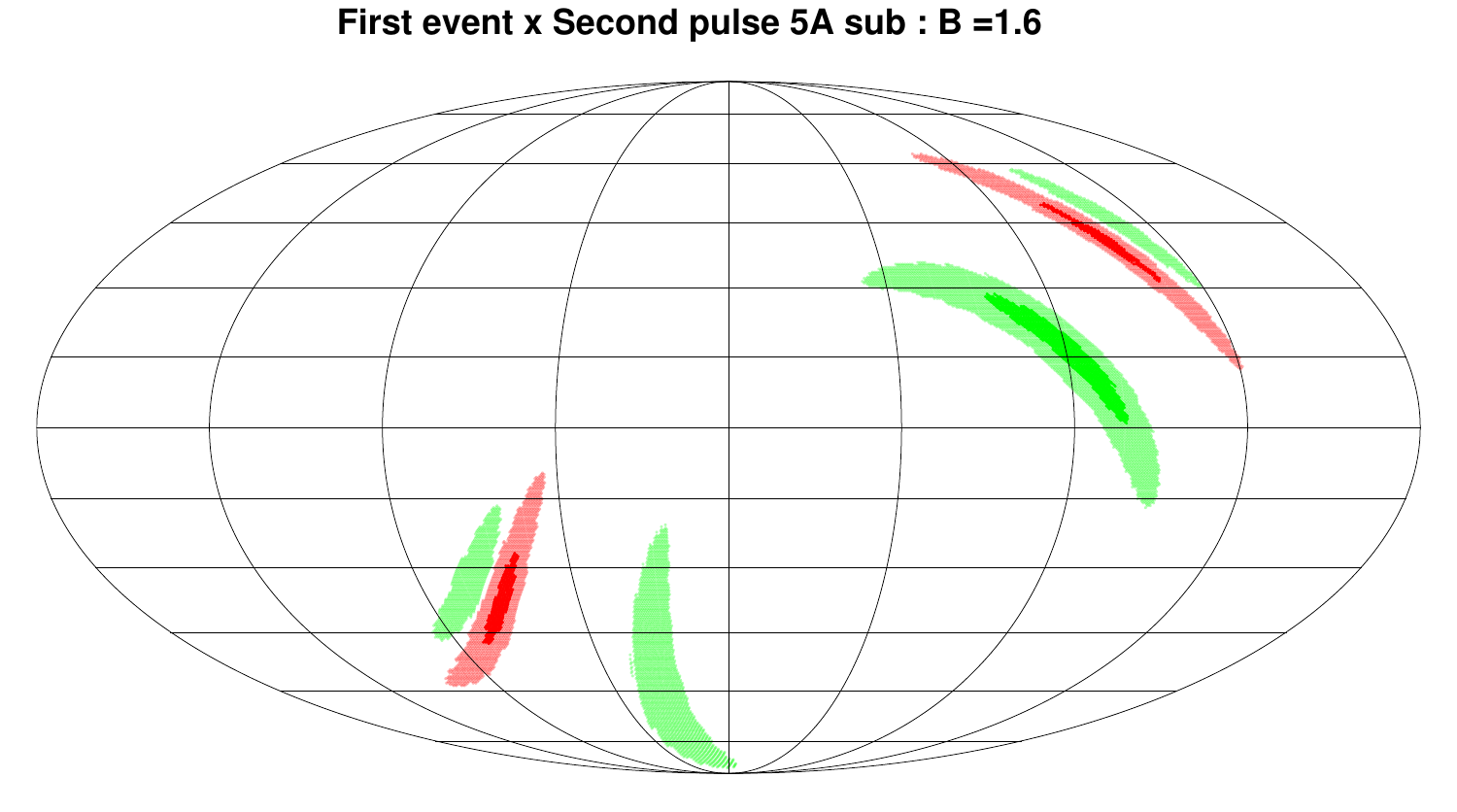}
    \includegraphics[width=0.35\textwidth]{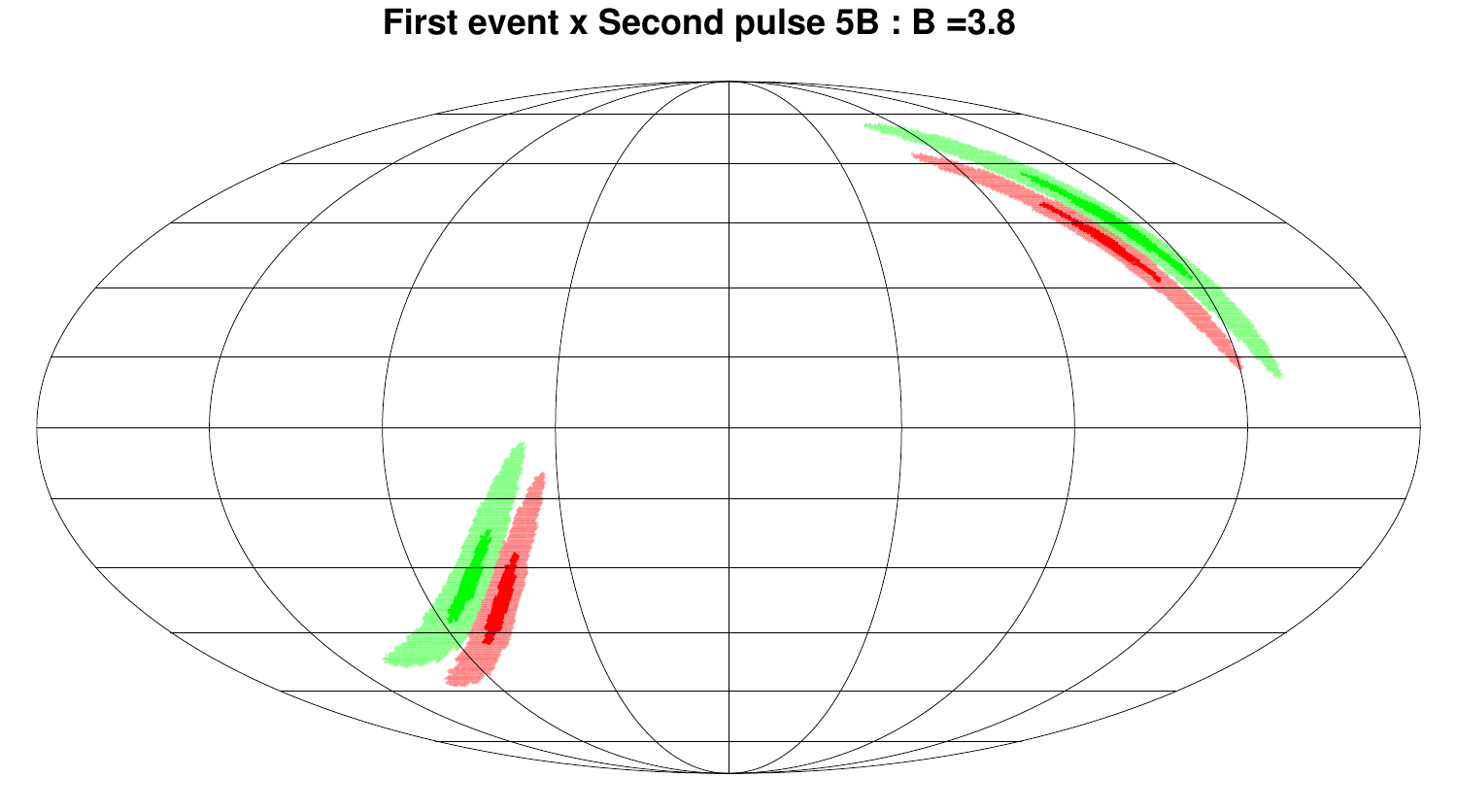}
    \includegraphics[width=0.35\textwidth]{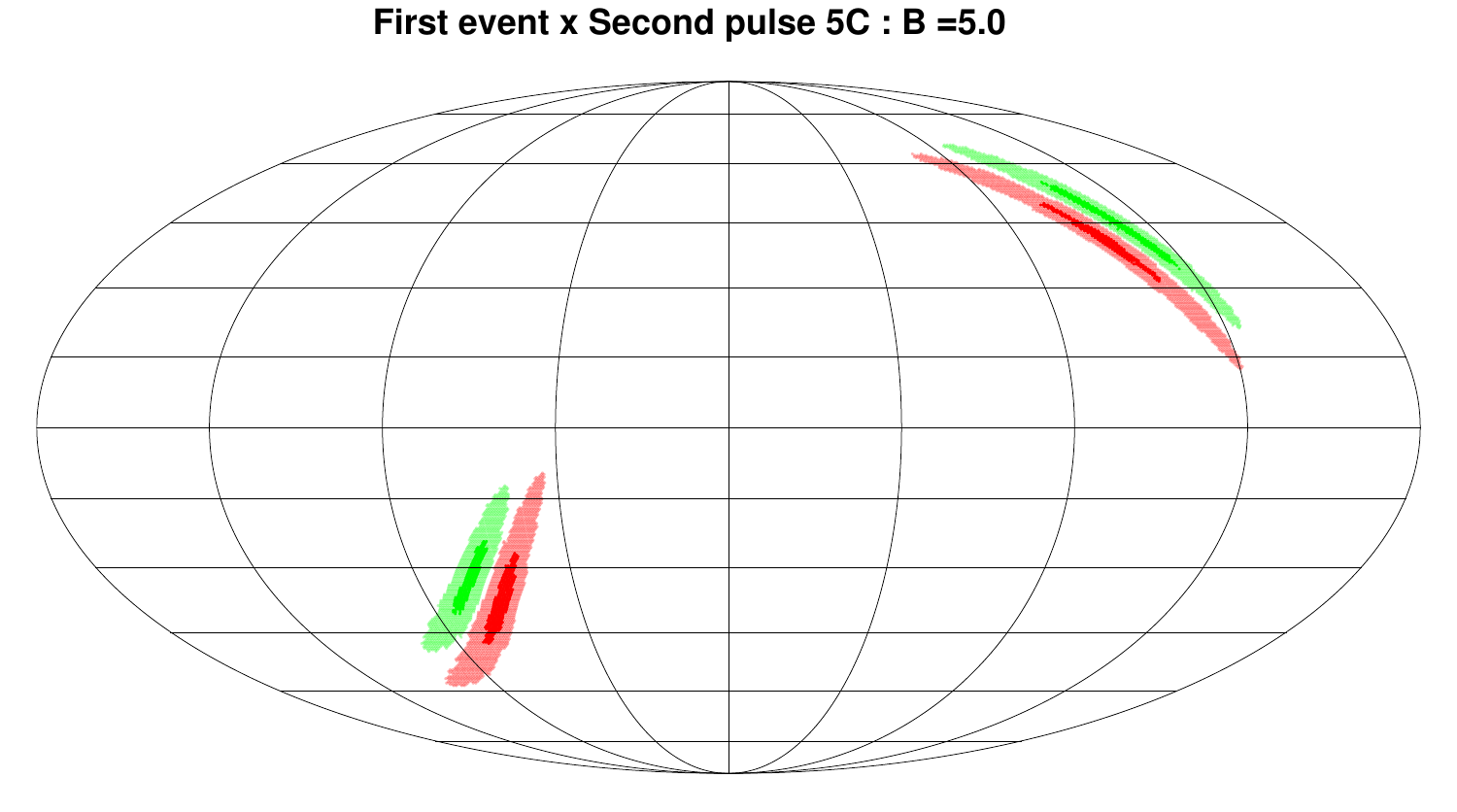}
    \includegraphics[width=0.35\textwidth]{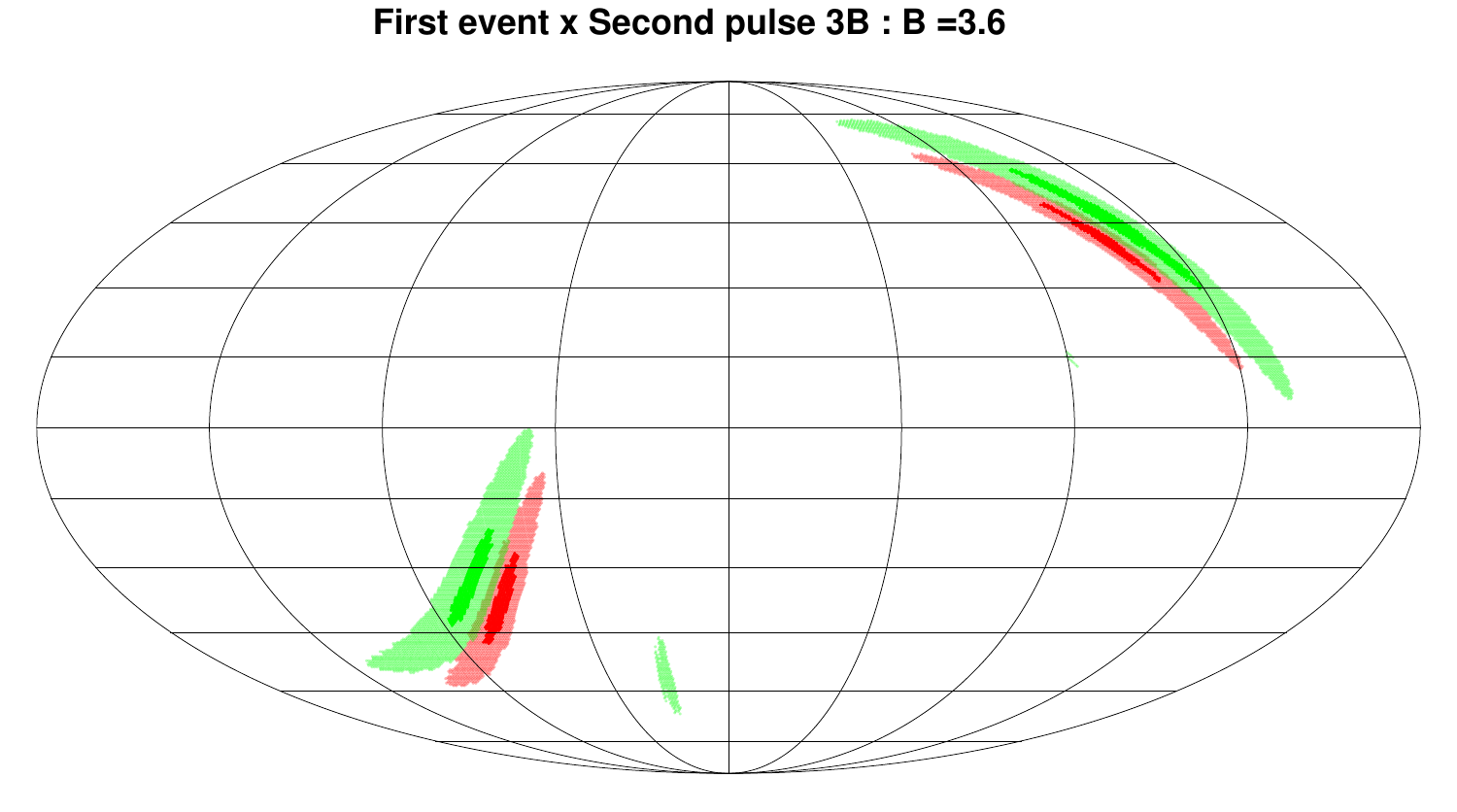}
    \includegraphics[width=0.35\textwidth]{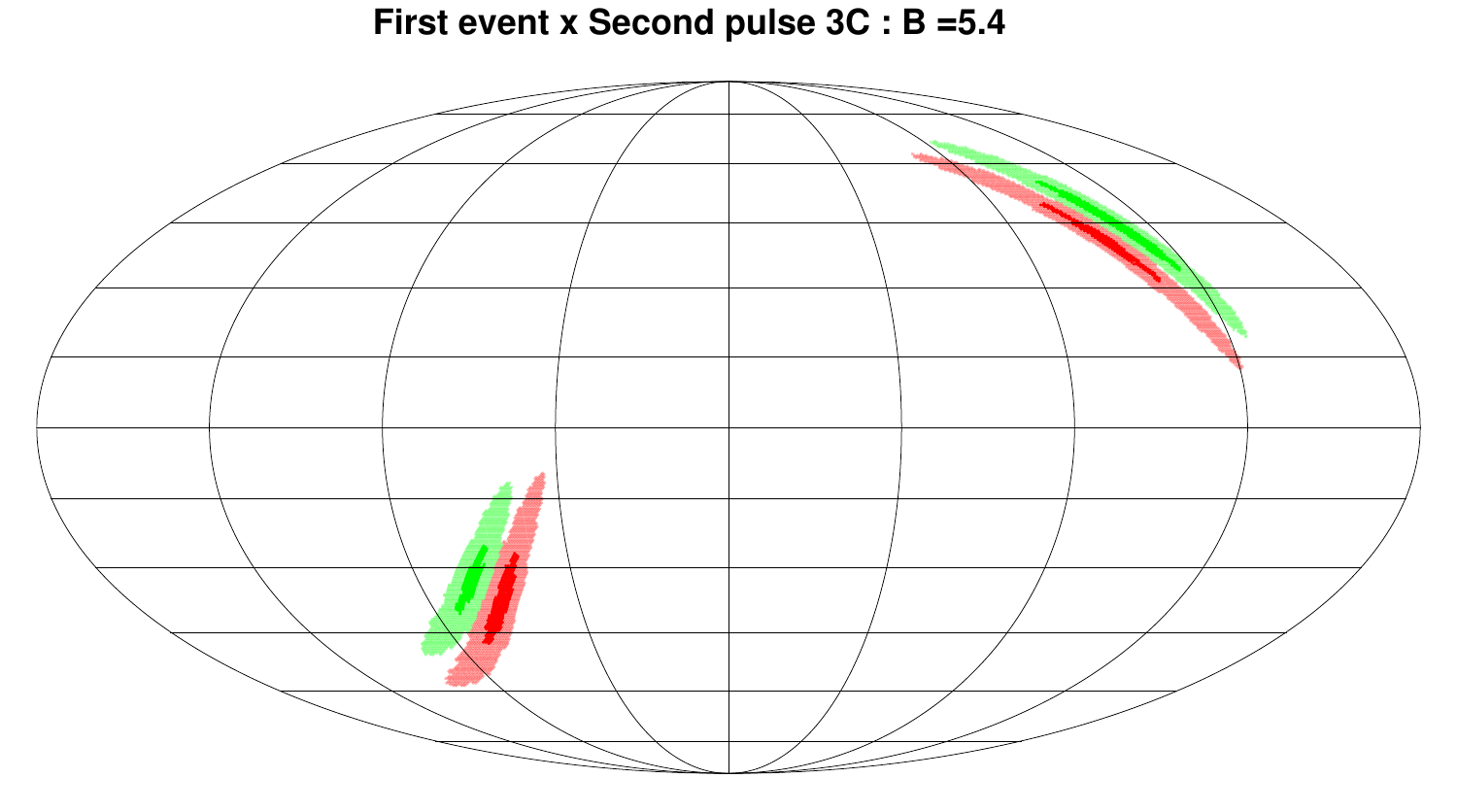}

    \caption{Co-localization analysis for GW151012. We performed several reconstructions with different search thresholds. The panels are named according to their thresholds: A-B-C index relates with different search parameters configurations and the numbers (5 or 3) relates pixel pattern configuration \cite{cWB}. All searches prefer the hypothesis of sky co-localization of echoes and main event, at Bayes factors of 1.6-5.4. }
    \label{GW151012}
\end{figure*}

We can first reproduce the GW151012 reconstruction performed in \cite{Salemi:2019uea} (see left panels of Fig. \ref{GW151012}). We imposed veto to the data (instead of subtraction) in order to exclude the main event, and performed several reconstructions with different search thresholds. The panels in Fig. \ref{GW151012} are named according to these thresholds: A-B-C index characterize different search parameter configurations, while the numbers (5 or 3) show the pixel pattern configuration \cite{cWB} (see \cite{GW190521echo} for details). The reconstructed waveform shown in the first panel of \ref{GW151012} correspond to the search 5A, with the co-localization analysis in the second panel of the same figure, which can be compared to Figs. 3 and 4 of \cite{Salemi:2019uea}). Other panels show the maps for different search thresholds. Using (\ref{eq:Bcoloc}), we see that all GW151012 searches prefer co-localization of echo and main event, at Bayes factors of $1.6-5.4$.

\section{Bayesian evidence while changing the waveform\label{appendixC}}
To validate that our method has correctly identified echo signals for GW190521, we also test it with a variety of different changes in waveform. As an example, Fig. \ref{Ringdown} represents posterior plot using the Ringdown waveform (including modes (l,m,n)=(2,2,0), (2,2,1) and (3,3,0)) instead of IMRPhenomXPHM and NRSur7dq4. Table \ref{table_appendix_1} demonstrates that the Bayesian evidence for stimulated Hawking radiation, is robust to the choice of template. In this table, $\delta_{t} h_{+/\times}$ implies that we include effect of change in polarisation via change in time delays between the metric polarizations $h_{+}$, $h_{\times}$ assuming either the strong (beyond GR) gravity regime does \cite{Ezquiaga:2020dao} change the polarisation of the post-merger signal. Time shift variable in this table accounts for the nonlinear dynamical effects changing the time delay of first echo. The sign variable is $\pm$ in Eq.\ref{eq:R}.

\begin{table*}[!ht]
\begin{center}
\begin{tabular}{ |c|c|c|c|c|c|c|c|c|c|c|c|c|c|c| }
\hline
$\mathcal{B}^{\rm{echoes}}$ & 7 & 4 & 3 & 7 & 4 & 4 & 4 & 5 & 8 & blue$9.1^{+2.7}_{-2.5}$ & 7 & 6 & $7.5^{+3.8}_{-2.3}$ & 6\\
\hline
waveform & IMR & IMR & IMR & R & R & R & IMR & IMR & IMR & R & R & IMR & NR & NR\\
\hline
$\phi$ & $\pi$ & $0$-$2\pi$ & $0$-$2\pi$ & $0$-$2\pi$ & $0$-$2\pi$ & $0$-$2\pi$ & $0$-$2\pi$ & $0$-$2\pi$ & $\pi$ & $0$-$2\pi$ & $0$-$2\pi$ & $0$-$2\pi$ & $0$-$2\pi$ & $0$-$2\pi$ \\
\hline
A & $0$-$1$ & $0$-$2$ & $0$-$2$ & $0$-$2$ & $0$-$2$ & $0$-$2$ & $0$-$1$ & $0$-$1$ & $0$-$1$ & $0$-$2$ & $0$-$2$ & $0$-$1$ & $0$-$2$ & 1 \\
\hline
sky location & unif & unif & fixed & unif & fixed & unif & unif & unif & unif & unif & unif & unif & unif & unif \\
\hline
n echoes & 3 & 2 & 2 & 3 & 3 & 3 & 3 & 3 & 3 & 3 & 3 & 3 & 2 & 2 \\
\hline
time shift & 0 & 0 & 0 & 0 & 0 & $\pm0.05$ sec & 0 & 0 & 0 & 0 & 0 & 0 & 0 & 0 \\
\hline
 $\delta_{t} h_{+/\times}$ & 0 & 0 & 0 & 0 & 0 & 0 & 0 & 0 & 0 & 0 & 0 & $\pm 7\%$ & 0 & 0 \\
\hline
sign in Eq.(\ref{eq:R}) & $+$ & $+$ & $+$ & $+$ & $+$ & $+$ & $+$ & $-$ & $-$ & $+$ & $-$ & $+$ & $+$ & $+$\\
\hline
$\log_{10} \Lambda$ & 0 & 0 & 0 & 0 & 0 & 0 & 0 & 0 & 0 & 0 & 0 & 0 & $0\pm13$ & $0\pm13$ \\
\hline
\end{tabular}
\caption{Bayes factor for echoes using different GR waveform family and/or echo model. Here IMR stands for IMRPhenomXPHM and R stands for (l,m,n)=(2,2,0), (2,2,1) and (3,3,0) ringdown model \cite{Capano:2021etf} and NR stands for NRSur7dq4. The fixed sky location values are chosen for the most likelihood of sky location RA= 3.5 rad (right ascension) and Dec=0.73 rad (declination). ``unif'' stands for uniform. }\label{table_appendix_1}
\end{center}
\begin{center}
\end{center}
\end{table*}

\section{cWB p-value estimation\label{SMB}}
In this section, we quantify the p-value of pixels in cWB using injections of GR waveform.

\begin{figure}
\centering
\includegraphics[width=0.5\textwidth]{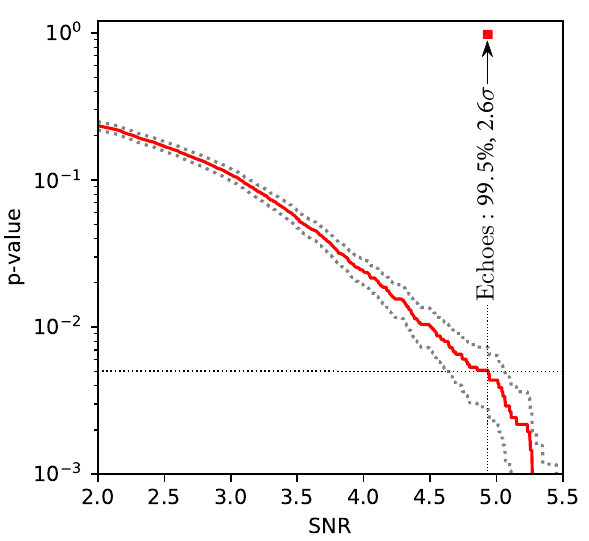}
\caption{cWB p-value estimation postmerger signal vs its signal-to-noise ratio and its Poisson error using injection approach. \label{pvalue}}
\end{figure}

In this approach we inject the maximum likelihood GR waveform in random places within $-32 \rm{\ sec}<t-t_{merger}<32 \rm{\ sec}$ excluding $-5 \rm{\ sec}<t-t_{merger}<5 \rm{\ sec}$ for a data for which the main event was already subtracted (to avoid any confusion for cWB search). In order to get better realizations of noise (off-source data) we also randomly time-shifted each detector data independently with maximum allowed shifts between -3 sec to +3 sec.
The search is implemented any time we inject the maximum likelihood GR waveform of the event for both user parameters of Figs. \ref{CWBpixels}a (up to 0.5 second post-merger) and  \ref{CWBpixels}b (up to 3 seconds post-merger). The comparison of SNR$^2$ of the two identified cWB triggers quantifies the significance of cWB echoes. As this difference is $24.3=15.98^2-15.20^2$ for the real event (Table \ref{Table_cwb}), we can estimate a p-value based on the number of times that $\Delta$SNR$^2$ exceeds this in random injections. Assuming that with almost equal chance backward and forward echoes may appear in  random noise, we exclude the backward pulses from our look-elsewhere search, yielding a p-value $= 5.1^{+2.2}_{-2.2}\times 10^{-3}$ (see Fig. \ref{pvalue}).

The attentive reader may note an apparent inconsistency between the Bayes factor found via PyCBC and the p-value computed via cWB. The reason behind this illusory paradox is that these numbers answer two very different questions. While the Bayes factor quantifies the Bayesian statistical support for the existence of echoes \textit{after} the GW190521 event for \textit{the specific} Boltzmann model, the aforementioned p-value gives us a frequentist  model-agnostic probability of finding a trigger with similar cWB significance to recovered echoes, from random noise, i.e. a measure of the ``look-elsewhere" effect. Therefore, there is no reason to expect them to represent same confidence levels.

\section{Cross correlation between detectors\label{xpectogram}}
As a complementary test to cWB and PyCBC, we use the method introduced in \cite{Abedi:2018npz} to correlate the two detectors. Here,  instead of Wiener filtering (which is more optimized for signals extended over LIGO bandwidth) we used whitening (dividing data by ASD rather than PSD), and do not add up multiple harmonics.
 Using the same method introduced in \cite{GW150914} we obtained the amplitude spectral density (ASD). Then we whitened the data by dividing by the ASD.
\begin{eqnarray}
&H(t,f)=\rm{Spectrogram}\left[ \rm{IFFT}\left( \frac{\rm{FFT}(h_{H}(t-\delta t))}{ASD_{H}} \right) \right],& \nn\\ 
&L(t,f)=\rm{Spectrogram}\left[ \rm{IFFT}\left( \frac{\rm{FFT}(h_{L}(t))}{ASD_{L}} \right) \right].& \nn\\ &&
\end{eqnarray}
We implemented the above function based on the same setup as LIGO code \cite{GW150914} to obtain the spectrogrm with mlab.specgram() function in Python with NFFT=$f_{s}=16384$, and mode='complex'.  Similar to \cite{GW150914}, the number of points of overlap between blocks is NOVL = $NFFT \times 15/16$. Note that with this setup spectrogram uses time $\pm 0.5$ sec in order to Fourier transform around each time.

The resulting spectrogram is cross-correlation of two detectors H1 and L1,
\begin{eqnarray}
X(t,f)= \Re\left[H(t,f) \times L^*(t,f)\right], \label{x_def}
\end{eqnarray}
which is the whitened cross-power spectrum of the two detectors. Taking into account the opposite phase of GW polarization for Hanford and Livingston, the real signals show up as peaks in $X(t,f)$.

\section{Effective time delay between Hanford and Livingston}\label{delta_t}
We find the time delay $\delta t$ between the two detectors via a naive estimation, explained in the following:

\begin{enumerate}
\item      One of the whitened strain series is shifted within (-10 msec, 10 msec).
\item      A complex spectrogram for both Hanford and Livingston detectors, $H(t,f)$ and $L(t,f)$ with 16k HZ data is obtained.
\item      Since the two detectors H1 and L1 have nearly 180 degrees phase difference, we maximized $|H(t-\delta t,f)-L(t,f)|$ for constant $f$ within the range $t-t_{\rm{merger}}=$(-2 s, +0.6 s) and $f=$(35 Hz, 100 Hz). 
\item     We only kept the 10 highest values of $|H(t-\delta t,f)-L(t,f)|_{f=f_{\rm max}(t)}$ and sum over them (we expect smaller values to be affected by noise).
\item      Finally, $\delta t$ is found having the maximized value of this sum, which happens at the value of $1.71$ ms.
\end{enumerate}

Note that, the time delay $\delta t = 1.71 $ ms is not a precise geometric time delay and is affected by small phase differences away from 180 degrees between the two detectors as well.
Interestingly, precise calculation with fix maximum likelihood sky location values of ra=3.5 rad and dec=0.73 rad corresponds to the detector time difference of 1.62 ms consistent with what has been obtained in this section.

Here Fig. \ref{Wider} represents result of this method in wider range in comparison to Fig. \ref{fig:sub-2}.

\begin{figure*}
\begin{minipage}[l]{0.46\textwidth}
\begin{subfigure}{0.9\textwidth}
  \centering
  \includegraphics[width=.9\linewidth]{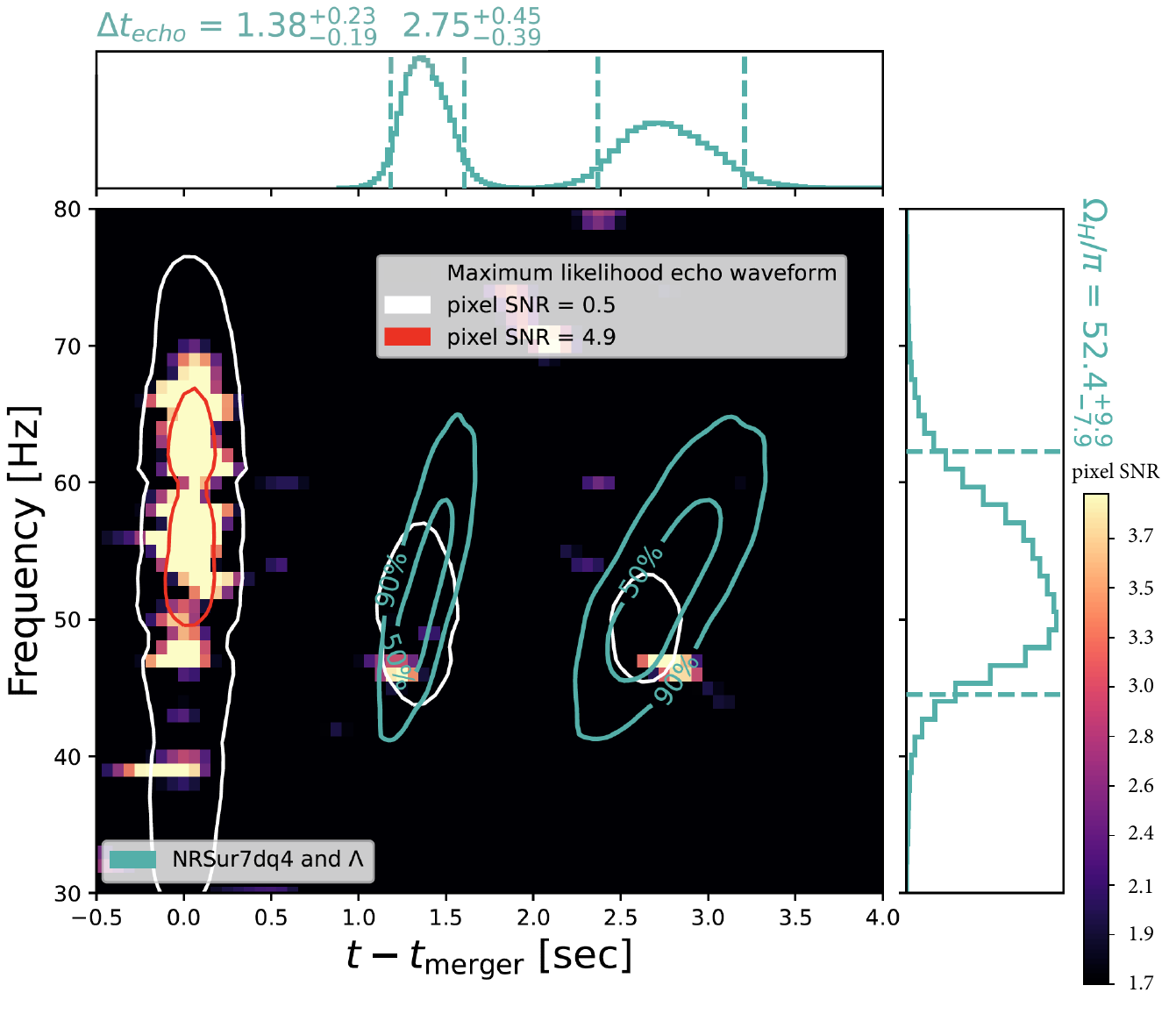}
  \caption{H1$\times$L1 whitened cross-spectrum (colors), the most likelihood PyCBC model (white and red contours), and the predicted super-Planckian echoes ($\Lambda=10^{5.5}$; light-sea-green)\label{Fig8b}.}
  \label{fig:sub-second}
\end{subfigure}
\end{minipage}
\hfill{}
\begin{minipage}[r]{0.43\textwidth}
\begin{subfigure}{1.0\textwidth}
  \centering
  \includegraphics[width=.9\linewidth]{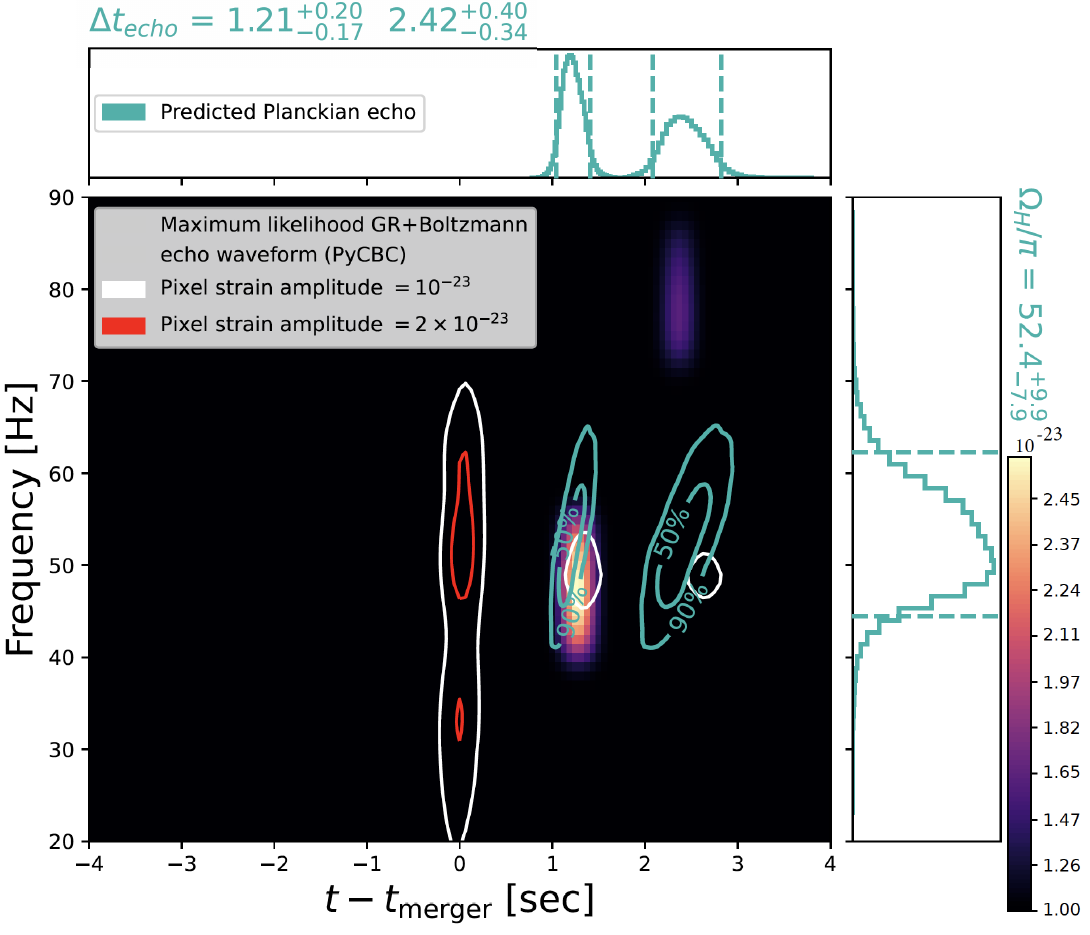}
  \caption{cWB reconstructed waveform (H1$\times$L1 [strain]$^2$) cross-spectrum using residual data compared to the maximum likelihood Boltzmann echo waveform (contours) using PyCBC assuming predicted Planckian echoes. In this plot we see that the 80 Hz feature may be affects the co-localization analysis.}
  \label{fig:sub-third}
\end{subfigure}
\end{minipage}
\caption{Comparison of three search methods: PyCBC, Cross-correlation, and cWB. White and red contours represent maximum likelihood Boltzmann echoes based on NRSur7dq4 general relativistic surrogate model. Contours with light-sea-green color represent 50\% and 90\% expected regions along with their 1D distributions, for both 1st and 2nd echoes.}
\label{Fig10}
\end{figure*}

\begin{figure*}
\centering
\includegraphics[width=0.8\textwidth]{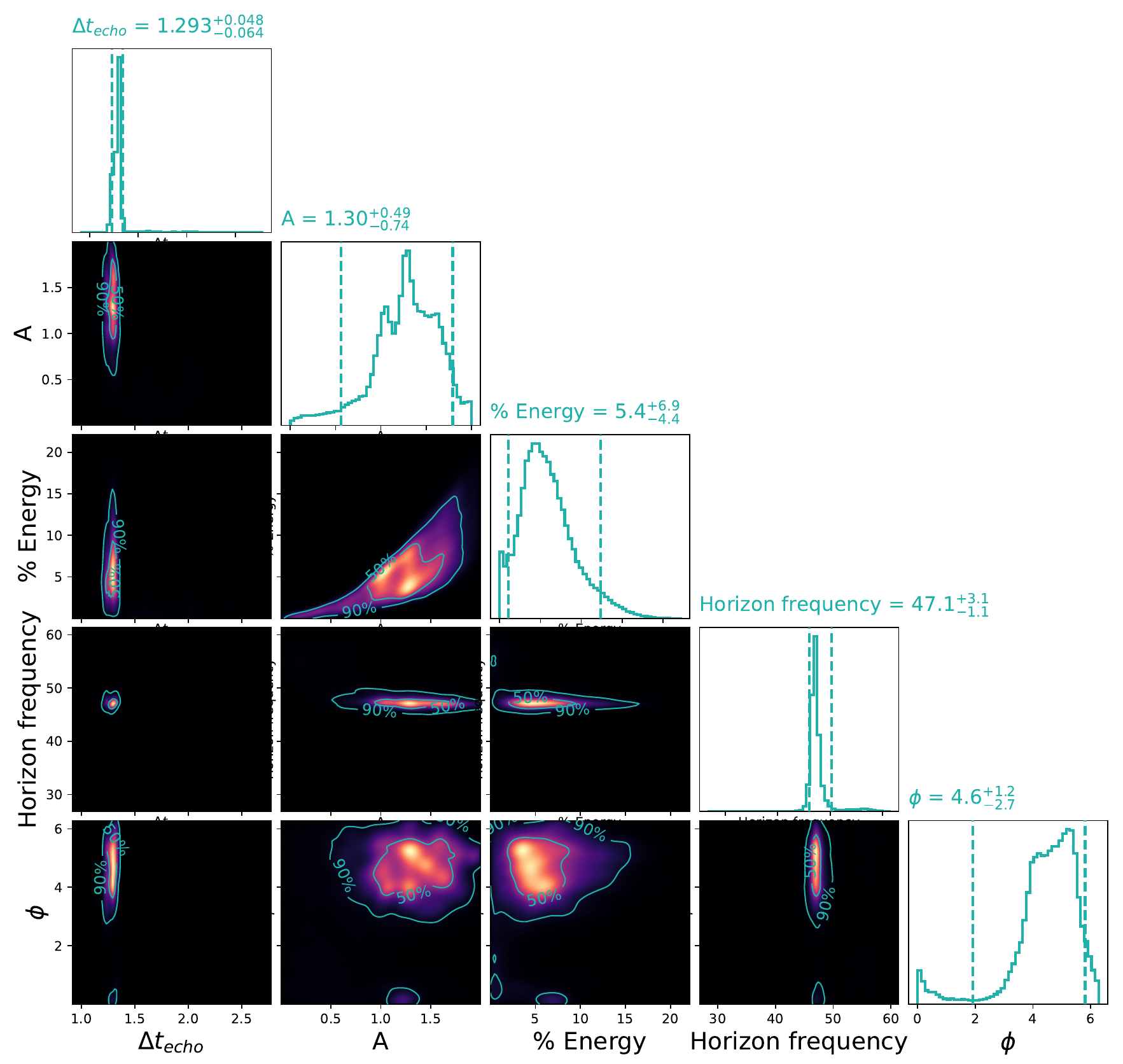}
\caption{Parameter estimation of Ringdown+(Boltzmann) Echoes waveform for GW190521. This plot has shown $\mathcal{B}^{\rm{Echoes}}=9.1^{+2.7}_{-2.5}$ in preference for Boltzmann echoes waveform. Contours in Off-diagonal plots indicate the 50\% and 90\% credible regions of 2D marginal posteriors. The diagonal plots show the 1D marginal posteriors, with the median and 90\% credible intervals where the recovered values are indicated at top of the plots.}
\label{Ringdown}
\end{figure*}

\begin{figure*}
    \begin{subfigure}{\textwidth}
    \centering
    \includegraphics[width=\textwidth]{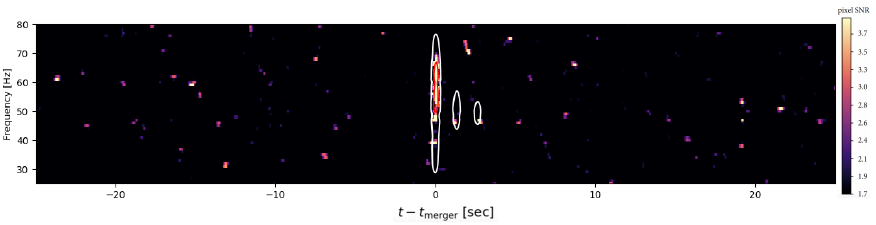}
    \end{subfigure}
    \begin{subfigure}{\textwidth}
    \centering
    \includegraphics[width=\textwidth]{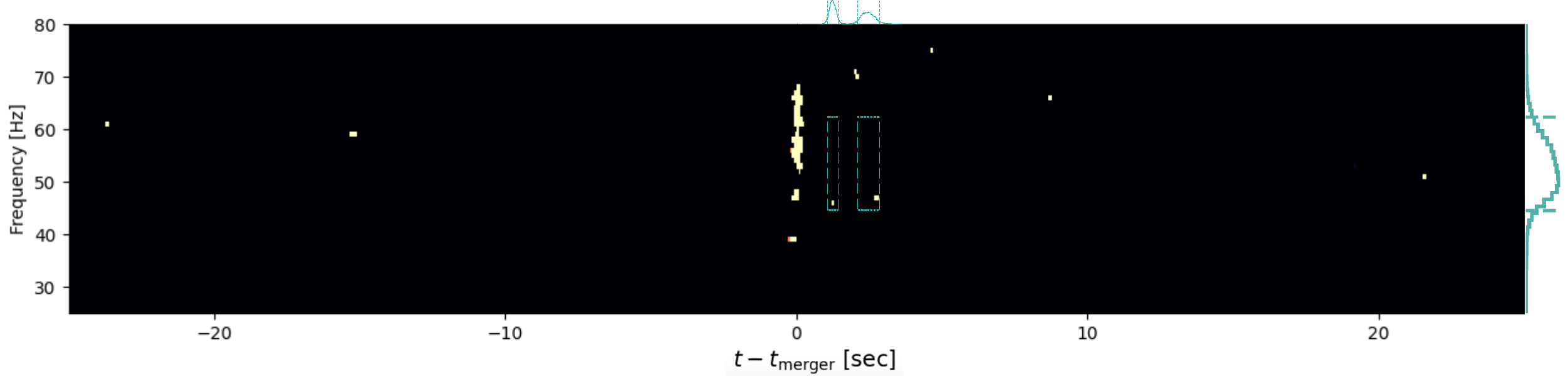}
    \end{subfigure}
    \caption{Look-elsewhere plot demonstrating that the echo signals found within Planckian 1st and 2nd echo contours are scarce in random noise. a) Same cross-correlation of H1 and L1 as Fig. \ref{fig:sub-2} within extended time of $-25\ \rm{sec}\leq t-t_{\rm{merger}}\leq +25\ \rm{sec}$ for GW190521 based on Eq.\ref{x_def}. b) Cross-correlation of H1 and L1 except for fixed pixel power corresponding the first echo pixel and 1D distributions show predicted 1st and 2nd Planckian echoes and horizon frequency.}
    \label{Wider}
\end{figure*}

\section{Comment for LVK about ADA waveform\label{ADA_method}}
Here, we perform our analysis using ADA phenomenological waveform \cite{Abedi:2016hgu}, with a physical prior on $\Delta t_{\rm echo}$.

Although the original ADA search was frequentist, based on SNR maximization and p-value estimation, we plug the waveform into PyCBC pipeline to perform a Bayesian evidence estimation.
Here, we examine the method with two different choices of priors.
\begin{figure*}
\begin{minipage}[l]{0.48\textwidth}
\begin{subfigure}{1.0\textwidth}
  \centering
  \includegraphics[width=1.0\linewidth]{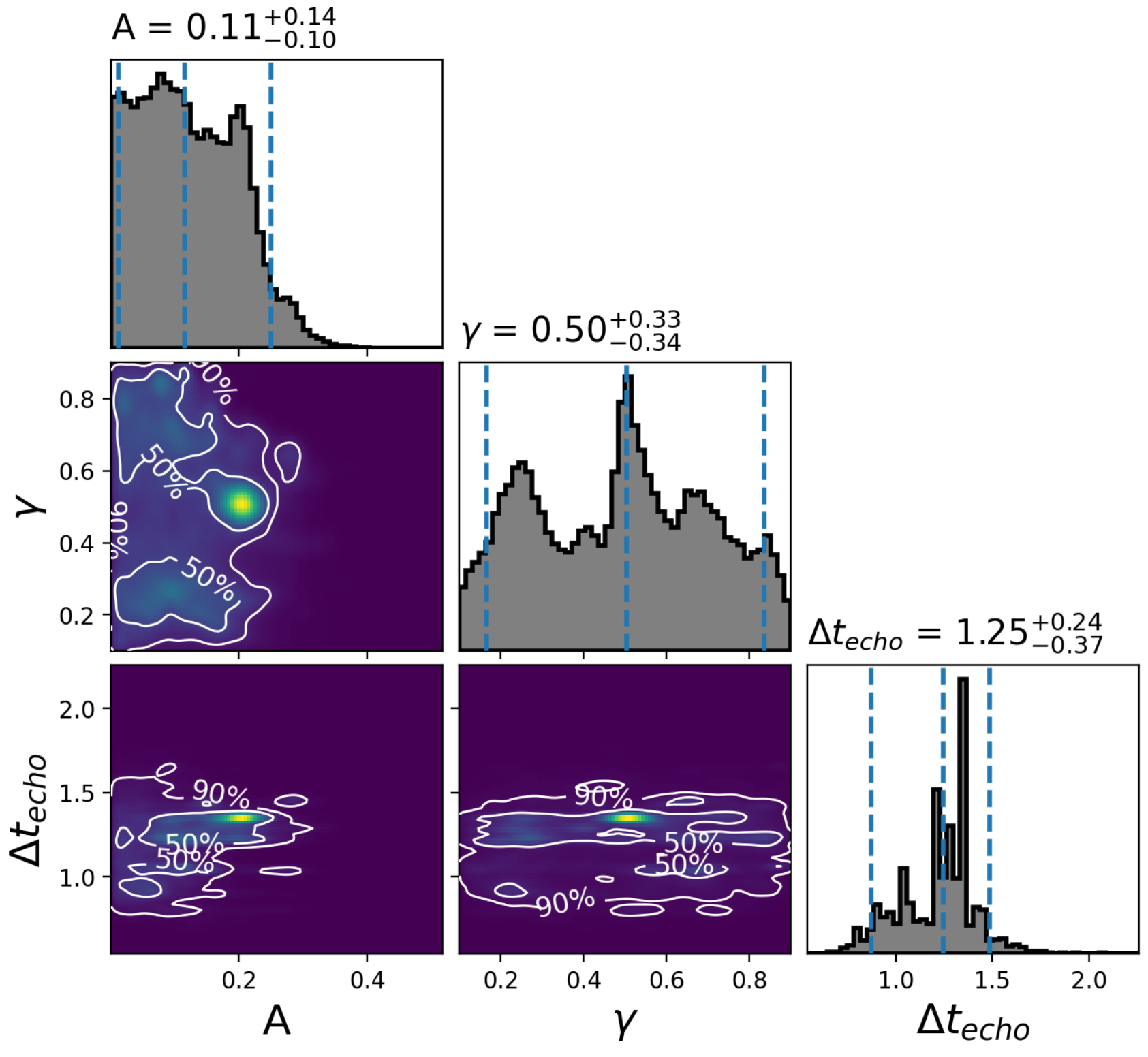}\hfill
  \caption{}
  \label{fig:sub-ADAa}
\end{subfigure}
\end{minipage}
\hfill{}
\begin{minipage}[r]{0.48\textwidth}
\begin{subfigure}{1.0\textwidth}
  \centering
  \includegraphics[width=1.0\linewidth]{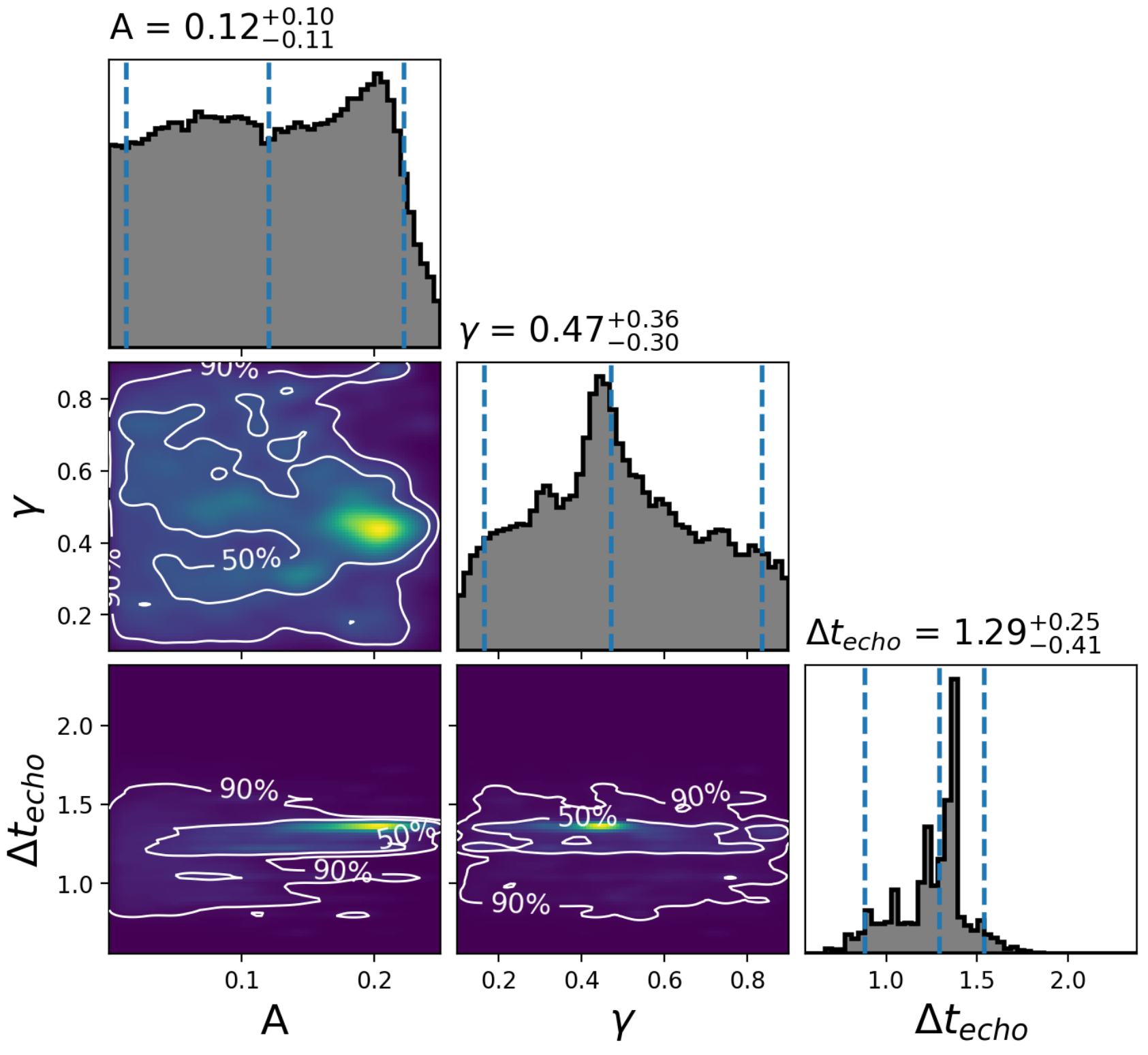}\hfill
  \caption{}
  \label{fig:sub-ADAb}
\end{subfigure}
\end{minipage}
\caption{(a,b): Two ADA \cite{Abedi:2016hgu} echo parameter estimation with PyCBC for GW190521.}
\label{FigADA}
\end{figure*}

The following choices are made for these two searches:
\begin{itemize}
\item Fig. \ref{fig:sub-ADAa}: We assume energy of echo to be less than energy of main event. We set the number of echoes$=10$. We also include $\Lambda$ as the parameter Planckian energy as what we used in this paper. 
\item Fig. \ref{fig:sub-ADAb}: We assume maximum amplitude if ADA wavform to be 0.25 and number of echoes$=2$. We also include $\Lambda$ as the parameter Planckian energy as what we used in this paper.
\end{itemize}
Note that the pipeline simultaneously fits for both GR and ADA echo parameters. Here we chose NRSur7dq4 as our base GR waveform.

The bayes factors of echoes for these two searches are 1.47 (Fig. \ref{fig:sub-ADAa}) and 0.52 (Fig. \ref{fig:sub-ADAb}) which are two orders of magnitude bigger than what was reported by LVK collaboration in \cite{Abbott:2020jks} (bayes factor $=1.5\times 10^{-2}$) using ADA for this event. This suggests that the prior used by the LVK search excluded the range with largest evidence.

\end{document}